\documentclass[reprint,eqsecnum,floats,aps,amsmath,amssymb%
,footinbib,prd,showpacs]{revtex4-1}
\usepackage{amsmath,amsthm,latexsym,amssymb,amsfonts}
\usepackage{enumerate}
\usepackage{graphicx}
\usepackage{subfigure}
\usepackage{calc}

\usepackage{LQC-symbols}

\input cyracc.def
\newfam\cyrfam
\font\tencyr=wncyr10
\font\sevencyr=wncyr7
\font\fivecyr=wncyr5

\textfont\cyrfam=\tencyr \scriptfont\cyrfam=\sevencyr
\scriptscriptfont\cyrfam=\fivecyr

\newcommand{\oq}{{\,{}^o\! q}} 
\newcommand{\cell}{\mathcal{V}} 
\newcommand{\Fou}{\mathcal{F}} 

\newcommand{\bR}{\boldsymbol{R}}

\begin{document}

\title{Prescriptions in Loop Quantum Cosmology: A comparative analysis}

\author{Guillermo A. \surname{Mena Marug\'an}${}^{1}$}
\email{mena@iem.cfmac.csic.es}
\author{Javier \surname{Olmedo}${}^{1}$}
\email{olmedo@iem.cfmac.csic.es}
\author{Tomasz \surname{Paw{\l}owski}${}^{2}$}
\email{tpawlows@unb.ca}

\affiliation{
  ${}^{1}$Instituto de Estructura de la Materia, IEM-CSIC,
  Serrano 121, 28006 Madrid, Spain
  \\
  ${}^{2}$Department of Mathematics and Statistics,
  University of New Brunswick, Fredericton, NB, Canada E3B 5A3.
}

\begin{abstract}
  Various prescriptions proposed in the literature to attain the
  polymeric quantization of a homogeneous and isotropic flat spacetime coupled
  to a massless scalar field are carefully analyzed in order to discuss their differences.
  A detailed numerical analysis confirms that, for states which are not deep in the quantum realm,
  the expectation values and dispersions of some natural observables
  of interest in cosmology are qualitatively the same for all the considered prescriptions.
  On the contrary, the amplitude of the
  wave functions of those states differs considerably at the bounce epoch for these prescriptions.
  This difference cannot be absorbed by a change of representation.
  Finally, the prescriptions with simpler superselection sectors are clearly more efficient
  from the numerical point of view.
\end{abstract}


\maketitle

\section{Introduction}

Loop Quantum Gravity (LQG) \cite{Thiemann-book,*Rovelli-book,al-status}, one of
the most promising approaches to unify general relativity with quantum
physics, has attracted a lot of attention in recent years. In particular, considerable
progress has been achieved in its application to symmetry reduced models for cosmology,
a field known as Loop Quantum Cosmology (LQC)
\cite{bojo-liv,*a-lqc-overview,*a-lqc-intro,*mm-lqc-overw,*mm-lqc2-overw}.
In this context, the analysis of the simplest (isotropic) cosmological systems
\cite{aps-det,aps-imp} has led to a qualitatively new picture of the early
universe dynamics \cite{aps-prl}, where the current expanding universe is
preceded by a --semiclassical \cite{cs-recall,kp-scatter}--
contracting one. This promising viewpoint opens new windows in modern
cosmology, resulting for example in a drastic increase of the probability for
inflation \cite{as-infl} and ensuring the geodesic completeness of the
(isotropic) cosmological spacetimes \cite{s-geodesic}. The original analysis has
been rigorously extended to various topologies \cite{apsv-spher,*van,*skl} and matter
contents \cite{bp-negL,kp-posL}, as well as to homogeneous but anisotropic cosmologies
\cite{mgp,*awe-b1,*mgp-ev}. Furthermore, recent years have witnessed growing
progress in the extension of the formalism to inhomogeneous scenarios,
particularly to Gowdy models (both in vacuo \cite{mgm-short,*mm,*mgm,*mmw} and with matter
\cite{mmm}) and to perturbative frameworks \cite{akl-qft}.
In addition, the generalization of the formalism to various Bianchi type models \cite{we-b9,*awe-b2}
has provided a viable hope for a general singularity resolution through the
Belinsky-Khalatnikov-Lifshitz mechanism \cite{as-bkl}. On the other hand, the reformulation
of LQG as a deparametrized theory \cite{gt-bk,*dgkl-qg} enables the application of the techniques of LQC
(either directly or after a suitable generalization) in the context of
the full theory, allowing one to check, in particular, the robustness of the LQC
results.

Despite the rapid advances and successful applications of LQC to systems of
increasing complexity, many of the basic aspects of the theory in its simplest
setting remain to be fully understood. One of them is the ambiguity inherent to the
construction of the quantum Hamiltonian constraint. The choice of
factor ordering (and densitization) gives rise to various quantization
prescriptions. While all of them provide the same physical picture
--to a high precision-- for the kind of states that are usually considered in
cosmology (admitting an epoch when the universe is semiclassical), any
possible physical or mathematical difference in other regimes must be investigated and discussed.

Here we address this question using as probe model
the simplest cosmological system with nontrivial evolution, i.e., a flat Friedmann-Robertson-Walker
(FRW) universe coupled to a massless scalar field. We focus our attention on four prescriptions,
three of which have already been studied in the literature. These are the original prescription
used in Ref. \cite{aps-imp} (known as \emph{APS}, from the initials of its authors), its corresponding simplification put forward in Ref.
\cite{acs}, which allows one to describe the dynamics analytically (known as \emph{sLQC}, which stands for {\sl solvable LQC}),
and the prescription of Ref. \cite{mmo-FRW} (denoted as \emph{MMO}, again from the initials of the authors). This latter prescription
is known to significantly simplify the
physical Hilbert space structure, asserts rigorously the generality of the bounce paradigm, and
leads to a unique Wheeler-DeWitt (WDW) limit in each of the superselection sectors that are (anti)symmetric under parity reflection.
The fourth prescription that we are going to analyze is a simpler version of this
\emph{MMO} prescription. In this article, we discuss the analytical and numerical implications
of the application of each of these prescriptions to construct the quantum model, and we
compare the details of the physical picture that they provide, investigating them in fully quantum
(not sharply peaked) states. In particular we will show that, while the expectation values
of certain natural observables show negligible discrepancies, physical differences
between prescriptions actually do exist, making them detectable, at least in principle.
This fact has important consequences for any effective or semiclassical treatment,
because it shows that the choice of representation and the details of the quantization
procedure actually \emph{can} have an imprint on the dynamics and have to be
taken into account.

This manuscript is organized as follows. In Sec.~\ref{sec:framework} we briefly
describe the classical system and the quantum framework. The prescriptions
analyzed in this article and their main properties are presented in
Sec.~\ref{sec:presc}. Our numerical methods and results are explained in
Sec.~\ref{sec:numerics} and Sec.~\ref{sec:results}. Finally,
Sec.~\ref{sec:concl} contains a general discussion and the main conclusions. In addition, an appendix
presenting the WDW quantum counterpart of the considered model is included.

\section{Classical and quantum framework}\label{sec:framework}

Let us first remind the construction and basic properties
of the model in LQC. The foundations and specifications of
this model have been discussed in Refs. \cite{abl-lqc,aps-imp}.
In particular, the APS prescription is described in Ref. \cite{aps-imp}.
Details about the other quantization prescriptions can be found in Refs. \cite{acs,mmo-FRW}.
We will briefly review them all, focusing on those steps where the prescriptions differ.

\subsection{The classical spacetime}\label{sec:frame-class}

The flat FRW spacetime admits an orthogonal foliation by spatial homogeneous $3$-surfaces
$\Sigma_t$ (parametrized by $t$). Its metric can be written in the form
\begin{equation}
  g = -N^2(t)\rd t^2 + a^2(t) \oq ,
\end{equation}
where $N$ is a lapse function, $a$ is the scale factor, and $\oq$ is a fiducial
Cartesian metric, constant in comoving coordinates.

The canonical description derived from the Einstein-Hilbert
action requires integrating the Lagrangian and Hamiltonian density over $\Sigma_t$.
To avoid divergences of the integrals when $\Sigma_t$ is noncompact,
one introduces an infrared regulator restricting the integration to a
cubical cell $\cell$ (again constant in comoving coordinates). The
geometrical degrees of freedom are coordinatized in the phase space by the
Ashtekar-Barbero connections and triads, which, owing to the isotropy of the system,
can be gauge fixed to the form
\begin{equation}
  A^i_a=cV_o^{-1/3}\delta_a^i , \quad  E^a_i=pV_o^{-2/3}\delta^a_i ,
\end{equation}
where $V_o$ is the volume of $\cell$ with respect to $\oq$. Then, all the information
about the geometry
is captured in the canonical pair $\{c,p\}= 8\pi \gamma G/3$ (where $\gamma$ is
the Immirzi parameter, fixed as explained in Ref. \cite{dl-gamma,*m-gamma}). The matter degrees
of freedom are described by the field $\phi$ and its canonical momentum
$p_\phi$, such that $\{\phi,p_{\phi}\}=1$.

The only nonvanishing constraint that remains after the gauge fixing is the
Hamiltonian one:
\begin{subequations}\label{eq:ham-class}\begin{align}
  \boldsymbol{C}(N) &= N (C_{\gr} + C_{\phi}), \label{eq:ham-lapse} \\
  C_{\gr} &= -\frac{6}{\gamma^{2}}c^2\sqrt{|p|} , \quad C_{\phi} = 8\pi G\frac{p^2_\phi}{|p|^{3/2}}.
\end{align}\end{subequations}
On shell $(C_{\gr} + C_{\phi}=0)$, it completely determines the dynamics of the system.

\subsection{Quantum foundations}\label{sec:frame-quant}

In order to quantize the system, we apply the Dirac program, first representing
the classical degrees of freedom as operators, momentarily ignoring the constraint ({\it kinematical level}).
The physical description is then obtained by imposing the constraint quantum mechanically.

\subsubsection{Kinematics}\label{ssec:kin}

The kinematical quantization is performed in two steps, each of them with a different approach.
For the matter content, we apply a standard Schr\"odinger
representation. The matter sector of the kinematical Hilbert space
is identified with $\Hil^{\phi}_{\kin}
=L^2(\re,\rd\phi)$, spanned by the basis of generalized eigenstates $(\phi|$ of the
field operator $\hat\phi$. The elementary operators are $\hat{\phi}$ (which
acts by multiplication in this representation) and $\hat{p}_\phi = -i\hbar\partial_\phi$.

In turn, the geometry is quantized adopting the methods of LQG (see Ref. \cite{abl-lqc}
for details). For an isotropic model, the standard holonomy-flux algebra can be
restricted to holonomies along straight edges and fluxes across unit squares with
respect to $\oq$. As a consequence, the configuration algebra $\textrm{Cyl}_S$ is an algebra
of almost periodic functions of $c$, and the kinematical Hilbert space for the gravitational sector
becomes $\Hil^{\gr}_{\kin} = L^2(\bar{\re},\rd\mu_B)$, where $\bar{\re}$ is the Bohr compactification
of the real line (with Bohr measure $\rd\mu_B$). A natural basis for $\Hil^{\gr}_{\kin}$
is formed by the eigenfunctions $|v\rangle$ of the triad operator $\hat{p}$
(which can be identified with the flux across a unit square)
such that $\hat{p}|v\rangle = \sgn(v)(2\pi\gamma\lPl^2\sqrt{\Delta}|v|)^{2/3}|v\rangle$, where
$\Delta$ is related with the
spectrum of the LQG area operator.
This basis is orthonormal: $\langle v | v' \rangle = \delta_{v,v'}$.

All geometric elements of the system (constraints, observables) can be expressed in terms
of two types of operators:
{i)} an oriented physical volume corresponding to the cell $\cell$, $V(\cell)=\sgn(p)|p|^{3/2}$, and {ii)} the holonomy
components $\N_{\mu}=e^{i\mu c/2}$, for an appropriate choice of $\mu$.
Actually, owing to the specifics of the quantization \cite{aps-imp}, the choice that must be adopted
in the construction of the Hamiltonian constraint is such that $\mu$ becomes a function of the triad,
$\bar{\mu}(p)$, and this function is fixed by the requirement that the
square loop with fiducial length
$\bar{\mu}$ built by the holonomies (in order to define the curvature)
has the minimum physical area that is allowed, $\Delta$. This choice corresponds to the so called \emph{improved dynamics} \cite{aps-imp}.
The action of these operators on the basis of $\Hil_{\kin}^{\gr}$ is:
\begin{equation}
  \hat{V}|v\rangle = \textrm{sgn}(v)|v|2 \pi \gamma l_{\textrm{Pl}}^2\sqrt{\Delta}|v\rangle , \quad
  \hat{\N}_{\bar{\mu}}|v\rangle=|v+1\rangle .
\end{equation}
Finally, the full kinematical Hilbert space is the tensor product
$\Hil_{\kin}=\Hil_{\kin}^{\gr}\otimes\Hil_{\kin}^{\phi}$.

\subsubsection{The quantum constraint}\label{ssec:constr}

We now express the constraint in terms of the operators introduced above.
This involves, in particular, an approximation to the curvature
using holonomies along a square loop of physical area equal to $\Delta$, as we have already
commented (see Ref. \cite{aps-imp} for the detailed procedure).
As a result, the Hamiltonian
constraint \eqref{eq:ham-class}, at a lapse function of reference, $N_0$, takes the general form
\begin{equation}\label{eq:constr-form1}
  \widehat{\boldsymbol{C}(N_0)} =  \widehat{N_0C_{\gr}} \otimes \id +\hat{B} \otimes \hat{p}_{\phi}^2 ,
\end{equation}
where $\hat{B}$ is some operator which is diagonal in the basis $\{|v\rangle \}$, and
$\widehat{N_0C_{\gr}}$ is a selfadjoint, difference operator of second order.
For both of them, the domain of definition is chosen to be
$\Cyl_S$.

Depending on the prescription, the operator $\hat{B}$ may involve the inverse volume,
which is again quantized via Thiemann methods and takes the form:
\begin{equation}\label{eq:inv-v}
  \begin{split}
    \left[\widehat{\frac{1}{V}}\right]
    &=\left(\frac{3}{4\pi\gamma l_{\textrm{Pl}}^2\sqrt{\Delta}}\right)^3\widehat{\text{sgn}(V)}
      |\hat{V}|  \\
    &\times \left(\hat{\mathcal N}_{-\bar{\mu}}|\hat{V}|^{1/3}\hat{\mathcal N}_{\bar{\mu}}
     - \hat{\mathcal N}_{\bar{\mu}}|\hat{V}|^{1/3}\hat{\mathcal N}_{-\bar{\mu}}\right)^3 .
  \end{split}
\end{equation}

In order to solve the constraint, it is convenient to bring it into the explicitly separable form
\begin{equation}\label{eq:constr-dens}
  \hat{\mathcal C} =8\pi G(- \hbar^2\hat{\Theta}\otimes\id + \id\otimes\hat{p}_\phi^2)
\end{equation}
through a process which is often called \emph{change of densitization} \footnote{The procedure was first applied
  in Ref. \cite{aps-det}, renouncing to a fully rigorous description owing to the lack of space.
  Its rigorous definition was first provided in Ref. \cite{mmo-FRW}, for the MMO prescription.
  For the APS prescription \cite{aps-imp}, the change of densitization was discussed
  in complete detail in Appendix A of Ref. \cite{klp-aspects}. The process of changing the densitization and the
  caveats related with its application were studied in the general setting in Ref. \cite{klp-ga}
}.

The resulting difference operator $\hat\Theta$ is a nonnegative selfadjoint \cite{kl-flat}
operator on $\Cyl_{S}$. For all the considered prescriptions, it is defined on $\Cyl_{S}$ as
\begin{equation}\label{eq:Theta}
  \hat{\Theta} = -\hat{\mathcal{N}}_{2\bar{\mu}}f(\hat{v})\hat{\mathcal{N}}_{2\bar{\mu}}
  -\hat{\mathcal{N}}_{-2\bar{\mu}}f(\hat{v})\hat{\mathcal{N}}_{-2\bar{\mu}}
  + f_o(\hat{v}) ,
\end{equation}
where $\hat{v}|v\rangle = v|v\rangle$ and
\begin{subequations}\label{eq:Theta-coeffs}\begin{align}
  f(v) &= \frac{3\pi G}{4} (v^2-2-\alpha) + O(v^{-2}) , \\
  f_o(v) &= \frac{3\pi G}{2} (v^2-\alpha) + O(v^{-2}) .
\end{align}\end{subequations}
Here, $\alpha$ is a constant whose value depends on the factor ordering chosen in
each prescription. The action of $\hat{\Theta}$ only relates states with support
on lattices or semilattices of step 4, denoted by $\mathcal{L}_\varepsilon$, where
$\varepsilon\in (0,4]$. Then, we can identify sectors $\Hil_{\varepsilon} \subset \Hil_{\kin}$
preserved under the action of $\hat{\Theta}$, as well as by all the observables considered in
this manuscript (see the definitions in Sec.~\ref{ssec:obs-def}). They form
superselection sectors.
At each of these sectors, the quantity $\varepsilon$ used as a label to characterize them,
can be thought of as proportional to the minimum physical volume.

By applying the analysis of Ref. \cite{kl-flat}, one can show that the restriction
of $\hat\Theta$ to each of the superselection sectors has essential and absolutely
continuous spectra that are both equal to $\re^{+}$. Its degeneracy depends both on $\varepsilon$ and on
the prescription, but in all the cases it is at most twofold.
The spectral decomposition of $\hat\Theta$ introduces a natural basis of generalized
eigenfunctions $(e^{\varepsilon}_k|$, solutions to
\begin{equation}\label{eq:e-def}
  [\hat\Theta e^{\varepsilon}_k](v) = 12\pi G k^2 e^\varepsilon_k(v),
\end{equation}
where $e^\varepsilon_k(v)=(e^\varepsilon_k|v\rangle$. The basis is normalized,
\begin{equation}\label{eq:e-norm}
  (e^{\varepsilon}_k|e^{\varepsilon}_{k'}) = \delta(k'-k) ,
\end{equation}
where the delta is defined on a domain $\bR$ which can be either $\re$ or $\re^{+}$,
depending on the degeneracy of the spectrum (see the discussion in
Sec.~\ref{sec:presc}).

Finally, let us clarify the physical meaning of the subleading coefficient
$\alpha$ in Eq. \eqref{eq:Theta-coeffs}. One can conveniently split the evolution operator into
\begin{equation}\label{eq:theta-split}
  \hat\Theta = \hat\Theta_o + \widehat{\delta\Theta} ,
\end{equation}
where $\widehat{\delta\Theta}$ is a compact operator that contains all the terms $O(v^{-2})$ neglected in Eq. \eqref{eq:Theta-coeffs}.
On the other hand, $\hat\Theta_o $ has quite a simple closed form when expressed in terms of $v$ and the corresponding canonical
momentum $b$ with $\{v,b\}=4$. Namely, introducing the transformation \footnote{In certain cases, the support
of the wave function is restricted to a proper subset of $\lat_{\varepsilon}$ (see Sec.~\ref{sec:presc}). In those situations, one extends
the function to $\lat_{\varepsilon}$ by setting $\psi(v)=0$ at the missing points.}
\begin{equation}
  [\Fou\psi](b) = \sum_{v\in\lat_{\varepsilon}} e^{ivb/4} \psi(v)
\end{equation}
and the coordinate change $x=\ln[\tan(b/4)]/2$, a simple computation yields \cite{kl-flat}
\begin{equation}\label{eq:theta-potential}
  \hat\Theta_o = -12\pi G \left[  \frac{\alpha+1}{4\cosh^2(2x)} + \partial_{x}^2  \right] .
\end{equation}
The prescription-dependent constant $\alpha$ acquires then a neat physical interpretation,
related to the potential strength (see Ref. \cite{kl-flat} for more details).

\subsubsection{Physical Hilbert space}\label{ssec:Hphys}

Taking into account that $\hat{\mathcal C}$
has an invariant dense domain in each $\mathcal{H}_{\varepsilon}$ and is essentially
selfadjoint there, one can construct the physical Hilbert space $\Hil_{\phy}$, for instance,
by applying \emph{group averaging} techniques \cite{m-gave1,*almmt-gave,klp-ga}
to Eq. \eqref{eq:constr-dens}. The result is
$\Hil_{\phy}=L^2(\bR,\rd k)$, where the domain $\bR$ is determined by
the degeneracy of the spectrum,
as explained above.
This result is completely equivalent to the deparametrization of the
constrained system by selecting $\phi$ as an internal time \cite{aps-det}. That procedure provides
two sectors, which correspond to positive and negative frequencies.
Without loss of generality, one can restrict all considerations e.g. to the positive frequency sector.
In addition, the deparametrization leads to a notion of dynamical evolution (consistent with the
Schr\"odinger evolution picture) given by the map
\begin{equation}\label{eq:evol-map}
  \re\ni\phi \mapsto \Psi_{\phi}(\cdot)=\Psi(\cdot,\phi);\quad \Psi\in\Hil_{\kin}^{gr} .
\end{equation}
The ``time translations'' are the unitary transformations
\begin{equation}\label{eq:evol-map-unit}
  \Psi_{\phi_o}(v)\mapsto\Psi_{\phi}(v)=e^{i\sqrt{|\hat\Theta|}(\phi-\phi_o)}\Psi_{\phi_o}(v),
\end{equation}
whose generator is the operator $\sqrt{|\hat\Theta|}$ (defined by the spectral decomposition
of $\hat{\Theta}$).

Finally, the lack of symmetry breaking interactions introduces into the system a large symmetry:
reflection of the triad orientation, $v\mapsto -v$. It is then possible to split the physical
space into two sectors: the symmetric and the antisymmetric ones. For the rest of our discussion, we
choose the symmetric sector. This choice does not affect the results.

The physical states take the form:
\begin{equation}\label{eq:phi-states}
  \Psi(v,\phi) = \int_{k\in\bR} \rd k \tilde{\Psi}(k) e^{\varepsilon}_k(v) e^{i\omega(k)\phi} ,
\end{equation}
where  $\omega(k) = \sqrt{12\pi G}|k|$, $e^{\varepsilon}_k(v)$ are the symmetric basis functions
\eqref{eq:e-def} corresponding to the superselection sector $\varepsilon$, and $\tilde{\Psi}\in\Hil_{\phy}$.

\subsubsection{Natural observables}\label{ssec:obs-def}

In order to extract the physics and test possible differences
between prescriptions, we need to introduce (a convenient set of) physical observables
on $\Hil_{\phy}$. The unitary mapping \eqref{eq:evol-map} allows us to promote any
well defined kinematical observable $\hat{O}$ to a physical one, $\hat{O}_{\phi}$, acting on
the wave function as follows
\begin{equation}\label{eq:Ophys-def}
  [\hat{O}_{\phi}\Psi](v,\phi) = e^{i\sqrt{|\hat\Theta|}(\phi-\phi_o)}
  \left.\left[\hat{O}\Psi(v,\phi)\right]\right|_{\phi_o} .
\end{equation}

For our analysis here, we select a set of observables that are frequently used
in cosmology: the function of the volume $\ln|\hat{v}|_{\phi}$
\footnote{This operator can be defined using the spectral resolution of $\hat{v}$ as far
  as zero is not in the discrete spectrum, something that happens in most of the cases analyzed here.
  For the remaining cases, i.e. when $v=0$ is included, one may still design
  ways to define the operator, as discussed in Appendix A of Ref. \cite{klp-aspects}.
}, the energy density $\hat{\rho}_{\phi}$ --obtained from
\begin{equation}\label{eq:rho-def}
  \hat{\rho} = \ \boldsymbol{:}\widehat{\frac{p_{\phi}^2}{2V^2}}\boldsymbol{:} \
  = \frac{\hbar^2}{2} \left[\widehat{\frac{1}{V}}\right] \hat\Theta \left[\widehat{\frac{1}{V}}\right],
\end{equation}
where the symbol ``$\boldsymbol{:}$'' stands for symmetric ordering--, and the Hubble parameter
$\hat{H}_{\phi}$, built from
\begin{equation}\label{eq:H-def}
  \hat{H} = -\frac{1}{4i\gamma\sqrt{\Delta}}(\hat{\mathcal{N}}_{4\bar{\mu}}-\hat{\mathcal{N}}_{-4\bar{\mu}}) .
\end{equation}
All of them leave the spaces $\Hil_{\varepsilon}$ invariant.

These operators will be analyzed in Sec.~\ref{sec:num-obs} where, for certain classes of states, we will investigate
the difference in their expectation values for the distinct prescriptions under study.
These differences between expectation values will be considered significant if they are at least of the order
of the dispersions of the corresponding observables.

\section{The prescriptions}\label{sec:presc}

Even at the classical level, there exists a freedom in defining the densitization of the Hamiltonian constraint
of the system. This amounts to identifying in Eq. \eqref{eq:ham-lapse} what is the density weight of the
function on phase space that provides
the constraint and which part simply plays the role of a Lagrange multiplier. Equivalently, one can define the densitization of
the constraint by providing a (nonvanishing) lapse of reference $N_0$, or at least specifying its density weight,
setting then the constraint equal to $\widehat{\boldsymbol{C}(N_0)}$, as in
Eq. \eqref{eq:constr-form1}. One can then regard $N/N_0$ as the associated Lagrange multiplier.
Although these considerations have no much relevance on
purely classical grounds,
different choices produce different expressions after rewriting the constraint in terms of triads and
holonomies; in particular, there appear slight changes in the holonomy dependence owing to the
regularization of the inverse volume terms (see Sec.~\ref{ssec:constr}).

These differences get even more important at the quantum level where, in addition to the already mentioned effects
of ``holonomization'',
we have the freedom to select a particular operator representation for the classical constraint,
including the choice of order in functions of noncommuting elementary operators
(commonly called the factor ordering ambiguity).

The above ambiguities have already given rise to several constructions for the geometric operator $\hat\Theta$.
Here, we will focus our attention on four of them, three of which have already appeared in the literature. They are the
\emph{Ashtekar--Paw{\l}owski--Singh} (APS) prescription \cite{aps-imp}, the \emph{solvable LQC} (sLQC) prescription \cite{acs},
and the \emph{Mart{\'i}n Benito--Mena Marug{\'a}n--Olmedo} (MMO) prescription \cite{mmo-FRW}. In addition to those three,
we will also consider a prescription
(that we call sMMO) that unifies some properties of the sLQC and MMO ones. These four prescriptions are described below, case by case,
pointing out the properties of the evolution operator and the form of the physical Hilbert space for each of them.

\subsection{APS}\label{sec:presc-APS}

This is the original prescription used in the pioneer analysis of the (improved) LQC dynamics \cite{aps-imp}
(see also Ref. \cite{klp-aspects} for a recent discussion of some aspects of this quantization).
The choices of density weight and operator representation are the following:
\begin{enumerate}[(i)]
  \item The densitization agrees with the standard one in full LQG.
  Formally, this amounts to setting $N_0=1$ in Eq. (\ref{eq:constr-form1}).
  \item For the constraint, one chooses the algebraically simple,
    symmetric factor ordering:
    \begin{equation}
      \widehat{N_0C_{\gr}} \propto -(\hat{\N}_{2\bar{\mu}}-\hat{\N}_{-2\bar{\mu}})\, \hat{V}\,
      (\hat{\N}_{2\bar{\mu}}-\hat{\N}_{-2\bar{\mu}}).
    \end{equation}
  \item In passing to the form \eqref{eq:constr-dens}, one requires that the emerging
    Schr\"odinger system is \emph{strictly} equivalent to the group averaging of
    the Hamiltonian constraint in its original form \eqref{eq:constr-form1}
    \footnote{See Appendix A of Ref. \cite{klp-aspects}}.
    \label{it:APS-sync}
\end{enumerate}

As a consequence of these settings, the operator $\hat{B}$ is given by
\begin{equation}
B(v) = \frac{27}{8} |v| \big| |v+1|^{1/3} - |v-1|^{1/3} \big|^3,\end{equation}
while,
the coefficients
of the evolution operator $\hat{\Theta}$ on the symmetric sector of the theory are
\begin{subequations}
\label{eq:Theta-APS-compl}\begin{align}
  f(v) &= [\beta(v+2)]^{1/2} \tilde{f}(v) [\beta(v-2)]^{1/2} , \\
  \begin{split}
    f_o(v) &= \beta(v) [(1-\delta_{v,-4})\tilde{f}(v+2) \\
           &\hphantom{=\beta(v)}\, + (1-\delta_{v,4})\tilde{f}(v-2) ]  ,
  \end{split}
\end{align}
\end{subequations}
where
\begin{subequations}\begin{align}
  \tilde{f}(v) &= \frac{3\pi G}{8} |v| \big| |v+1| - |v-1| \big| , \\
  \beta(v) &= \begin{cases}
                [B(v)]^{-1} , & v\neq 0 , \\
                0 ,           & v=0 .
             \end{cases}
\end{align}\end{subequations}
Similar expressions are obtained in the antisymmetric sector (except for the $\delta_{v,\pm 4}$ terms, which do not contribute in that case).

Furthermore, the implementation of \eqref{it:APS-sync} imposes on the physical wave
functions \eqref{eq:phi-states} the constraint
\begin{equation}\label{eq:phys-0-constr}
  \Psi(0,\phi) = 0 .
\end{equation}

Therefore, in the precise form presented here for this prescription,
the state $|v=0\rangle$ \emph{does not decouple
from the rest of the domain} from the start (note that in the original representation of Ref. \cite{aps-imp}
the wave function still has a nontrivial value at $v=0$ \cite{klp-aspects}).
However, the apparent nondecoupling is only formal inasmuch as the state $|v=0\rangle$
\emph{does not contribute} to the space of physical states anyway.

The constant $\alpha$ describing the potential term
in Eq. \eqref{eq:theta-potential} takes the value:
\begin{equation}
  \alpha = \alpha_{\APS} = \frac{5}{9} .
\end{equation}

The structure of the physical Hilbert space depends on the superselection sector. We distinguish two cases:
\begin{enumerate}[(i)]
  \item For $\varepsilon=0,2$ (in the case $\varepsilon=0$, upon our symmetry/antisymmetry assumption)
    the subdomains $v>0$ and $v<0$ decouple, and the eigenspaces of the (extended) evolution operator
    are nondegenerate. As a consequence, the physical Hilbert space is $\Hil^{\varepsilon=0,2}_{\phy} = L^2(\re^+,\rd k)$.
    This case is called \emph{exceptional} (following Ref. \cite{aps-det}).
  \item When $\varepsilon\neq 0,2$, the two triad orientations are interconnected and the eigenspaces of $\hat\Theta$ are
    two-dimensional. The resulting physical Hilbert space is $\Hil^{\varepsilon}_{\phy} = L^2(\re,\rd k)$. We will further
    refer to this case as the \emph{generic} one.
\end{enumerate}

This dichotomy affects, in particular, the form of the symmetric superselection sectors of the theory. For exceptional cases,
one can restrict all considerations to functions supported on semilattices
\begin{equation}\label{eq:sub-semilat}
  \lat^\pm_{\varepsilon} = \{\pm (\varepsilon + 4n),n\in \natu \}
\end{equation}
and next extend them to fully symmetric domains by parity reflection.

For generic cases the superselection sectors have support on full lattices, of the form
$\lat_{\varepsilon}=\{\varepsilon+4n,n\in\integ\}$. These lattices are not invariant with respect
to reflection, and therefore one has to work with \emph{the union of two lattices},
$\lat_{\varepsilon}\cup\lat_{4-\varepsilon}$, first constructing the state
on $\lat_{\varepsilon}$ and then extending it by parity. This will force us to use different techniques
in Sec.~\ref{sec:num-basis} when identifying
the basis of symmetric functions $e_k^\varepsilon(v)$ numerically.

One of the unfortunate properties of the generic sectors in this prescription is the fact that the physical Hilbert space
is twice larger than for exceptional sectors. For the physically interesting applications (analysis of the universes which are
semiclassical and expanding at late times), one restricts the study in practice to half of the physical space, thus spanned by
only half of the basis functions. As we will see in  Sec.~\ref{sec:num-basis}, one chooses them to resemble
as close as possible (in certain aspects) the basis of the exceptional cases. This is achieved by imposing
additional requirements on the behavior of their geometrodynamical limit (the selection procedure and numerical
construction will be presented in the mentioned section). The remaining basis elements can be
then defined as the orthogonal completion of the constructed subset.

\subsection{sLQC}\label{sec:presc-sLQC}

This prescription, first suggested in Ref. \cite{acs}, has been proposed to bring
the evolution operator into a form as simple as possible, so that
the study of the system can be carried out in a fully analytic way.
The density weight and operator representation of the constraint are as follows:
\begin{enumerate}[(i)]
  \item The constraint is defined with density weight equal to one. Formally, this corresponds to a choice of the type $N_0=V/(8\pi G)$,
  for which the time parameter is synchronized with the scalar field.
  \item The gravitational part of the constraint is defined with the ordering
    \begin{equation}
      \widehat{N_0C_{\gr}} \propto - \hat{V} [ \hat{\N}_{2\bar{\mu}} - \hat{\N}_{-2\bar{\mu}} ]^2 \,\hat{V} .
    \end{equation}
\end{enumerate}
With the above criteria, in particular, the operator $\hat{B}$ in Eq. \eqref{eq:constr-form1} is just the identity; so no change of densitization
is needed in the quantum theory to reformulate the constraint and attain separation of the geometric and matter variables.
The resulting coefficients of the evolution operator are thus much
simpler than for the APS prescription and read
\begin{subequations}\label{eq:Theta-sLQC}\begin{align}
  f(v) &= \frac{3\pi G}{4} \sqrt{|v+2|}|v|\sqrt{|v-2|} , \\
  f_o(v) &= \frac{3\pi G}{2} v^2 ,
\end{align}\end{subequations}
which implies that the subleading term in Eq. \eqref{eq:Theta-coeffs} is
\begin{equation}
  \alpha = \alpha_{\sLQC} = 0 .
\end{equation}
One of the consequences of expression \eqref{eq:Theta-sLQC} is the annihilation of the state
$|v=0\rangle$ by $\hat\Theta$. The zero volume state therefore decouples from the evolution.
One can then superselect this state by its own and remove the quantum counterpart of the
singularity from the start. The mechanism of singularity resolution is thus
slightly different from that of the APS prescription.

The form of $\hat\Theta$ simplifies also the construction of the energy density operator,
which in this case is
\begin{equation}
  \hat{\rho} = - \frac{3}{32\pi G\Delta\gamma^2}(\hat{\N}_{2\bar{\mu}}-\hat{\N}_{-2\bar{\mu}})^2 \ .
\end{equation}
It is then possible to show that the \emph{entire spectrum} of $\hat{\rho}$, and not just its essential
part \cite{klp-aspects} (as in the case of other prescriptions), is the interval
$[0,\rho_c]$, where $\rho_c$ is the critical energy density \cite{aps-imp}.

Another convenient property of this prescription is the fact that, although in the
representation used in Eq. \eqref{eq:theta-potential} the operator $\hat\Theta$ still includes
a nontrivial potential (apart from the compact remnant), there exists an equivalent representation
in which it adopts a simple Klein-Gordon form $\hat\Theta = -\partial_{\bar{x}}^2$
(see Eqs.~(3.16) and (3.17) of Ref.~\cite{acs}).

The structure of the physical Hilbert space is exactly the same as for the APS prescription.
Again one has two cases: $(i)$ the exceptional case for $\varepsilon=0,2$, where the triad
orientations either decouple or the degeneracy is removed by the parity
symmetry restriction, and $(ii)$ the generic case, for all other values of $\varepsilon$.
The treatment is exactly the same
as in the APS prescription; in particular, the physical Hilbert space is
$\Hil^{\varepsilon=0,2}_{\phy}=L^2(\re^+,\rd k)$ in the exceptional cases,
and the eigenspaces are nondegenerate then, whereas for the generic case one has
$\Hil^{\varepsilon}_{\phy}=L^2(\re,\rd k)$ and a twofold degeneracy
(although in practice we will restrict the study to just half of $\Hil^{\varepsilon}_{\phy}$,
given the kind of semiclassical states that we want to consider).

\subsection{MMO}\label{sec:presc-MMO}

This prescription was originally motivated by the analysis of Bianchi I cosmologies
in the LQC scenario \cite{mgm-short,*mm,*mgp}, which give rise to natural proposals that,
when applied to the isotropic situation, affect
the factor ordering in the representation of the constraint. The main difference
with respect to the previous prescriptions is the appearance in
$\widehat{\boldsymbol{C}(N_0)}$ of an operator $\widehat{\sgn(v)}$. In the isotropic context, the prescription
was first introduced and studied in Ref. \cite{mmo-FRW}, following the analysis of its anisotropic
counterpart. It is characterized by
\begin{enumerate}[(i)]
  \item The density weight is the same as in LQG. Formally, the lapse can be viewed as $N_0=1$.
  \item The gravitational part of the constraint is
    defined as the reduction of its analog in the Bianchi I model \cite{mgp}, by
    identifying the degrees of freedom corresponding to distinct eigendirections
    of the metric. \label{it:B1red}
  \item This and the presence of $\sgn(v)$ allows one to choose the following operator representation for the constraint:
    \begin{equation}\label{eq:MMO-gen-form}
      \widehat{N_0C_{\gr}} = -\left[ \hat{A}
        \boldsymbol{:} \left(\hat{\N}_{2\bar\mu}-\hat{\N}_{-2\bar\mu}\right) \widehat{\sgn(v)} \boldsymbol{:}
        \hat{A} \right]^2 ,
    \end{equation}
    where $\hat{A}$ is certain operator that is diagonal in $|v\rangle$ and satisfies $\hat{A}|0\rangle=0$.
    We have used the notation $[\boldsymbol{:}\hat{X}\hat{Y}\boldsymbol{:}]\,=(1/2)[\hat{X}\hat{Y}+\hat{X}\hat{Y}]$.
\end{enumerate}
In order to achieve separation of variables, one can change the densitization and reformulate the constraint
by a procedure that is directly
inherited from the Bianchi I model. Namely, one can first deal with
the non-isotropic Bianchi I constraint and then reduce the result to the isotropic case. In this way,
one gets a constraint of the form \eqref{eq:constr-dens}, with coefficients for the evolution operator
\eqref{eq:Theta} that are given by
\begin{subequations}\label{eq:Theta-MMO}\begin{align}
  f(v) &= \frac{\pi G}{12} g(v+2)g(v-2)g^2(v)s_+(v)s_-(v) , \\
  \begin{split}
    f_o(v) &=  \frac{\pi G}{12} g^2(v)\big\{[g(v+2)s_+(v)]^2 \\
    &\hspace{1.9cm} +[g(v-2)s_-(v)]^2\big\},
  \end{split}
\end{align}\end{subequations}
where
\begin{subequations}\begin{align}
  \label{s}
    s_\pm(v) &= \sgn(v\pm2)+\sgn(v), \\
  \label{g}
    g(v) &= \begin{cases}
	      \left|\left|1+\frac1{v}\right|^{\frac1{3}}
		-\left|1-\frac1{v}\right| ^{\frac1{3}}
		\right|^{-\frac1{2}} &  v\neq 0, \vspace*{.2cm}\\
	      0 &  v=0.\\
	    \end{cases}
\end{align}\end{subequations}

The resulting evolution operator has several interesting properties:
\begin{enumerate}[(i)]
  \item The operator is an explicit square, $\hat{\Theta}\propto\hat{\Omega}^2$, where
    $\hat{\Omega}$ is a known second-order difference operator (see Eq.~(7) of \cite{mmo-FRW}).
  \item The coefficients $f(v)$ and $f_o(v)$ vanish in the whole interval $v\in[-2,2]$.
\end{enumerate}
The latter implies, in particular, that the states whose support corresponds to different
orientations of the triad are not mixed under the action of the constraint. Furthermore,
one can see that the superselection sectors corresponding to $\hat{\Theta}$ now have support
on semilattices \eqref{eq:sub-semilat}, and the absolutely continuous spectrum
of $\hat{\Theta}$ in each of them
is positive and nondegenerate. Hence, \emph{each} superselection sector (without any exception)
has the structure found for the exceptional situations in the other prescriptions presented above.

This in turn implies that the exact structure of $\Hil_{\phy}$ not only becomes the same for \emph{all}
the superselection sectors (in contrast with the previous prescriptions), but also that it is technically
simpler to deal with it (as we will discuss in detail in Sec.~\ref{sec:numerics}).

Another interesting property follows directly from the nondegeneracy of the spectrum. Namely, all the
basis elements converge in the limit of large $v$ to WDW \emph{exact} standing waves. This property,
which holds always in this prescription, is achieved in the previous two prescriptions \emph{only}
for the exceptional cases $\varepsilon=0,2$. For the remaining sectors, the discussed limit presents a small
(decaying exponentially with $k$) but nonvanishing deviation from the standing waves,
analogous to the case of a tunneling through a potential barrier.

Finally, the asymptotic expansions of $f(v)$ and $f_o(v)$ for $v\to\infty$ give
\begin{equation}
  \alpha = \alpha_{\MMO} = \frac{5}{3} .
\end{equation}

\subsection{sMMO}\label{sec:presc-sMMO}

In this prescription, which has not been stated explicitly in the literature so far,
the density weight and the operator representation of the constraint are selected to bring
together the nice features of the sLQC and MMO prescriptions:
\begin{enumerate}[(i)]
  \item The density weight is chosen as in the sLQC prescription, so that one directly attains
  separation of the geometric and matter variables in the constraint.
  \item The constraint operator is defined as a reduction of its Bianchi I counterpart,
    as in the MMO prescription.
\end{enumerate}
Again, the presence of $\sgn(v)$ and the parallelism with Bianchi I allow one to choose
an operator ordering such that the gravitational part of the constraint is of the form
specified in Eq. \eqref{eq:MMO-gen-form} (but with the operator $\hat{A}$ differing from
that of the MMO prescription). With these choices, $\hat{\Theta}$ is given by the
following coefficients:
\begin{subequations}\label{eq:Theta-sMMO}
\begin{align}
  f(v) &= \frac{3\pi G}{16} \sqrt{|v+2|}|v|\sqrt{|v-2|}s_+(v)s_-(v) , \\
  f_o(v) &=  \frac{3\pi G}{16} |v|\left[|v+2|s_+^2(v) + |v-2|s_-^2(v)\right] ,
\end{align}
\end{subequations}
with $s_\pm(v)$ defined in Eq. \eqref{s}.
The asymptotic behavior of these coefficients leads to a potential term in Eq. \eqref{eq:theta-potential}
with
\begin{equation}
  \alpha = \alpha_{\sMMO} = 0 .
\end{equation}

This prescription shares all the qualitative properties of the MMO one: the spectrum, the decoupling of
the $|v=0\rangle$ state, the decoupling of the triad orientations for all superselection sectors,
as well as all the properties following from this (see Sec.~\ref{sec:presc-MMO}).
On the other hand, the resulting evolution operator differs from the one of the sLQC prescription only by a
diagonal operator supported on $v\in (-4,4)\setminus\{0\}$. For the sector $\varepsilon=0$ that difference vanishes.
This implies that the system is exactly solvable for $\varepsilon=0$, while for the remaining sectors the difference
is just a tiny correction (see the comparison in Sec.~\ref{sec:results}), which for a noncompact system vanishes in the
limit in which the regulator is removed.

\subsection{Measuring the differences between prescriptions}\label{sec:presc-diff}

In principle, the considered prescriptions may lead to distinct physical pictures.
To analyze this possibility, we will investigate in Sec.~\ref{sec:numerics} the behavior of the set of standard cosmological
observables introduced above. But, independently of the existence of
significant differences for these observables and the feasibility of its detection, it is clear that
there are discrepancies in the physical sector of the theory and that these cannot be absorbed
just by a change of representation. In fact, we observe differences both in the exact structure of $\Hil_{\phy}$ and in the
(subleading) potential term of $\hat{\Theta}$, characterized by the constant $\alpha$ [see Eq. \eqref{eq:theta-potential}].

This implies that there indeed exist observables which can detect the differences between prescriptions,
even though they may be not of the greatest interest from a physical viewpoint. One observable of this kind is the
densitized constraint $\hat{\mathcal{C}}$ itself. Actually, given a prescription $\boldsymbol{A}$ corresponding to either APS, sLQC, MMO, or sMMO,
the constraint $\mathcal{\hat C}_{\boldsymbol{A}}$ will obviously annihilate all physical states
$|\Psi_{\boldsymbol{A}}\rangle\in\Hil_{\phy}^{\boldsymbol{A}}$, while it will not do so generically for physical states
of any other prescription $\boldsymbol{B}\neq\boldsymbol{A}$ (provided the action of $\hat{\mathcal{C}}_{\boldsymbol{A}}$
can be defined on them).
Mathematically, this means that
\begin{equation}
    \forall \boldsymbol{A}\neq\boldsymbol{B},\ \begin{cases}
      \forall \chi\in\Hil_{\kin}\,:\
        &(\Psi_{\boldsymbol{A}}|\hat{\mathcal{C}}_{\boldsymbol{A}}|\chi\rangle = 0 , \\
      \exists \chi\in\Hil_{\kin}\,:\
        &(\Psi_{\boldsymbol{A}}|\hat{\mathcal{C}}_{\boldsymbol{B}}|\chi\rangle \neq 0 .
    \end{cases}
\end{equation}
The difference between two constraints is of the form
\begin{equation}\label{eq:DeltaTheta}
  \hat{\mathcal{C}}_{\boldsymbol{A}}-\hat{\mathcal{C}}_{\boldsymbol{B}} = \widehat{\Delta\Theta}_{\boldsymbol{A}\boldsymbol{B}}\otimes\id , \quad
  \widehat{\Delta\Theta}_{\boldsymbol{A}\boldsymbol{B}} = \hat{\Theta}_{\boldsymbol{A}} - \hat{\Theta}_{\boldsymbol{B}}  ,
\end{equation}
where $\widehat{\Delta\Theta}_{\boldsymbol{A}\boldsymbol{B}}$ is a well defined kinematical observable in $\Hil_{\kin}^{\gr}$.
We can then define a family of physical observables $\widehat{\Delta\Theta}_{\boldsymbol{A}\boldsymbol{B}}|_{\phi}$ following the procedure explained in
Sec.~\ref{ssec:obs-def}. Such family allows us not only to detect the differences between prescriptions,
but also to pinpoint its variation in the evolution.

To understand the nature of these differences, let us note that the operator $\widehat{\Delta\Theta}_{\boldsymbol{A}\boldsymbol{B}}$
can be split like in Eq. \eqref{eq:theta-split}, which gives in the (momentum) $b$ representation:
\begin{equation}\label{eq:DTheta-split}
  \widehat{\Delta\Theta}_{\boldsymbol{A}\boldsymbol{B}} = 3\pi G \frac{\alpha_{\boldsymbol{B}}-\alpha_{\boldsymbol{A}}}{\cosh^2(2x)}
  + \widehat{\delta\Theta}_{\boldsymbol{A}}-\widehat{\delta\Theta}_{\boldsymbol{B}} .
\end{equation}
The compact term $\widehat{\Delta\delta\Theta}_{\boldsymbol{A}\boldsymbol{B}}=\widehat{\delta\Theta}_{\boldsymbol{A}}-\widehat{\delta\Theta}_{\boldsymbol{B}}$
can be neglected in the limit in which the infrared regulator is removed.
The only residual difference between the prescriptions is thus generated
by the potential term in Eq. \eqref{eq:DTheta-split}. Owing to the shape of the potential, the maximum difference is expected to
occur near the bounce point $x=0$. Its global effect can be understood physically as a slight difference in the dispersion of the
free Klein-Gordon wave packets \footnote{These are the wave packets in the coordinate $x$ defined in Sec.~\ref{ssec:constr}. They correspond
to bouncing solutions by themselves, and are distinct from both the WDW wave packets of Appendix~\ref{app:WDW} and
the wave packets of the sLQC prescription defined in Ref. \cite{acs}.} by this potential.

\section{Numerical analysis}\label{sec:numerics}

The different properties of the studied prescriptions force us to apply different numerical methods in our analysis.
The elementary ``bricks'' from which we construct the physical states are (some of) the eigenfunctions
of the operator $\hat\Theta$, which form an orthonormal basis of the gravitational part of the kinematical Hilbert space.
We will thus start (in Sec.~\ref{sec:num-basis}) with a detailed
explanation of the procedure to build this basis. We will then describe the procedures to
construct the physical states and evaluate the expectation values and the dispersions of the observables. This will be done
in Sec.~\ref{sec:num-states} and \ref{sec:num-obs}, respectively.
Finally, in Sec.~\ref{sec:effic} we will discuss and compare the efficiency and precision of the
methods used in different regimes.
The details
of most of the numerics employed in our analysis can be found in Refs.~\cite{aps-det,aps-imp,mmo-FRW}. Here, we will concentrate mainly on aspects
that are new or have not been sufficiently explored before. Our starting point is provided by appendices B and A.2
of Refs.~\cite{aps-det} and \cite{kp-scatter}, respectively.

\subsection{Basis construction}\label{sec:num-basis}

As we have already commented, the operator $\hat{\Theta}$ has
a continuous spectrum, and is thus diagonalizable in a basis of generalized
eigenfunctions [solutions to Eq. \eqref{eq:e-def} with an infinite kinematical norm].
Depending on the degeneracy of the spectrum, these basis elements are supported on
semilattices (nondegenerate case) or entire lattices (degenerate case), and can be
determined following different procedures.

\subsubsection{Nondegenerate eigenfunctions}\label{ssec:eig-semi}

This is the simplest situation from a technical point of view.
It occurs in the exceptional superselection sectors ($\varepsilon=0,2$)
of the APS and sLQC prescriptions and in all sectors of the MMO and sMMO ones.
In all these cases, the particular form of the evolution operator $\hat\Theta$ is such that
the eigenfunctions $(e^\varepsilon_{k}|$ (where $k\in\re^{+}$) are uniquely determined
by their initial value $e^\varepsilon_{k}(\varepsilon)$. To fix the phase of the eigenfunctions,
we choose this initial piece of data to be positive (see Refs. \cite{aps-imp,mmo-FRW}). Once the values at
$v>0$ are obtained, the function is extended to $v<0$ by symmetry requirements:
$e^\varepsilon_{k}(-v)=e^\varepsilon_{k}(v)$.

Let us now explain how to get $e^\varepsilon_{k}(v)$ for $v>0$.
One can see \cite{kp-scatter} that the asymptotic limit $v\to \infty$ of
$e^\varepsilon_{k}(v)$ has the form
\begin{equation}\label{wdw-limit}
  e^{\varepsilon}_{k}(v) \to r [
  e^{i\phi^{\varepsilon}_k}\, \underline{e}_{k}(v) +
  e^{-i\phi^{\varepsilon}_k}\, \underline{e}_{-k}(v) ],
\end{equation}
where $\phi^{\varepsilon}_k$ is a phase shift, $r$ is a positive real number, and
$\underline{e}_{\pm k}(v)$ are the generalized eigenfunctions of the
WDW analog of $\hat{\Theta}$ (see Appendix \ref{app:WDW}).
Given that $\underline{e}_{\pm k}(v)$ are (delta-)normalized as in Eq. \eqref{wdw:eig-norm},
the normalization condition \eqref{eq:e-norm} implies that $r=2$ (see appendix
A.2 of Ref.~\cite{kp-scatter}). The relation between the initial value $e^\varepsilon_{k}(\varepsilon)$
and the normalization factor $r$ is a priori unknown and can be determined only once the
eigenfunction (its $v\to\infty$ limit) is evaluated.
To overcome this problem, we divide the evaluation into several steps:
\begin{enumerate}[(i)]
  \item evaluation of a non-normalized eigenfunction $\psi_k$;
  \item finding its WDW limit to determine its norm $\|\psi_k\|$ relative to the condition \eqref{eq:e-norm};
  \item rescaling the eigenfunction to reach a normalized one:
    $\psi_k(v) \mapsto e^{\varepsilon}_k(v) = \|\psi_k\|^{-1}\psi_k(v)$.
\end{enumerate}

In the first step, we construct $\psi_k$ by setting
$\psi_k(\varepsilon) = 1$. The eigenfunction is then evaluated using
Eq. \eqref{eq:e-def} in an iterative process, point by point in the domain $\lat^+_{\varepsilon}\cap I$, where
$I=[\varepsilon,v_M]$. The boundary $v_M\gg k$ is chosen to lie (whenever technically possible)
deeply in the regime where the corrections to the asymptotic behavior \eqref{wdw-limit}
are small. In the present simulations, following numerical estimations, we fix
\begin{equation}\label{eq:vM}
  v_M \approx 4\cdot \min[10^7,\max\{100\cdot k,\exp(3\pi/k)\}] \ .
\end{equation}
This choice ensures also that, for small $k$, the selected interval contains
at least one oscillation period.

The second step is completed by a method analogous to the transfer matrix technique used in the
proof of Eq. \eqref{wdw-limit} in Ref. \cite{kp-scatter}. Namely, the value of the eigenfunction at
each pair of consecutive points in its domain of definition is represented as a linear combination of
the WDW basis functions, adopting the form
\begin{equation} \label{stand-wave}
	\psi_{k}(v) = \sqrt{\frac{2}{\pi v}}\, \tilde{r}_k(v)\cos[kx+\phi_k(v)],
\end{equation}
where $x=\ln v$, and the $v$-dependent coefficients $\tilde{r}_k(v)$ and $\phi_k(v)$
converge according to Eq. \eqref{wdw-limit} to their respective limits $\tilde{r}^{\epsilon}_k$ and
$\phi_k^{\varepsilon}$, respectively, with a rate
\begin{subequations}\label{eq:amp-phase}\begin{align}
  \tilde{r}_k(v) &= \tilde{r}^{\epsilon}_k \left[1+O\left(\frac{k^2}{v^2}\right)\right] , \label{eq:amp-phase-r} \\
  \phi_k(v) &= \phi_k^{\varepsilon}\left[1+O\left(\frac{k^2}{v^2}\right)\right] .
\end{align}\end{subequations}
The normalization is now determined by the identity $\tilde{r}_k^{\varepsilon}=2||\psi_{k}||$.
The limit  $\tilde{r}_k^{\varepsilon}$ can be evaluated quite easily.
For that we only need a sequence of points $\{(v_n^{-1},\tilde{r}_k^n)\}$
with $n\in\natu$, extrapolating numerically the desired limit at $v^{-1}\to 0$
(see appendix B of Ref.~\cite{aps-det}). In practice, we choose the sequence of points $v_n$ to
approximately follow the behavior $v_n\approx 2^{-n} v_0$, and use a polynomial
extrapolation (Neville's method). The specific method to evaluate $\tilde{r}_k^n$
depends on the value of $k$, namely:
\begin{enumerate}
  \item If $k\, x_M>2\pi$ (large $k$), with $x_M=\ln v_M$, the WDW limit (being
    a standing wave) has a wavelength small enough as to oscillate at least a few times in the
    chosen domain. Collecting the information at the extrema of these oscillations, we build a set of pairs
    $\{(v_n^{-1},\tilde{r}_k^n)\}$.
    The precise algorithm to evaluate these pairs is the following:
    \begin{enumerate}
      \item we find an extremum of $\psi_k(v)$, namely, a point
        $v_n\in\lat^+_\varepsilon \cap I$ where
        $|\psi_k(v_n)|>|\psi_k(v_n+4)|$ and $|\psi_k(v_n)|>|\psi_k(v_n-4)|$; initially, we look for the extremum
        closest to $x_M$;
      \item we extend $\psi_k$ to the interval $[v_n-4,v_n+4]$ via a polynomial
        interpolation of second-order (in $x$); the resulting function has the form
        \eqref{stand-wave} up to fourth-order corrections;
      \item given this interpolating function, we determine the pair
        $\{(v_n^{-1},\tilde{r}_k^n)\}$ corresponding to its extremum;
      \item we repeat the procedure, searching for
        the next extremum close to the point $x_n - \ln2$; the procedure is repeated
        iteratively until we obtain a sequence of five points, or we enter the region
	where the loop corrections become significant.
    \end{enumerate}
  \item If $k\,x_M<2\pi$ (small $k$), the wavelength of the oscillations is larger
    than $x_M$, and we do not get a sufficient number of extrema in the selected
    domain. Therefore, we modify the procedure explained above as follows:
    for each value of $v_n$, with $x_n = \ln v_n$,
    instead of searching for an extremum, we consider the pair of consecutive
    points $(v_n,v_n+4)$ and solve algebraically Eq. \eqref{stand-wave} to find
    $(r_k(v),\phi_k(v)$).
\end{enumerate}
The procedure used for small values of $k$ is simpler, since it does not involve
identifying extrema nor interpolating, but is less accurate than the one employed for large $k$'s.

Once the sequence $\{(v_n^{-1},\tilde{r}_k^n)\}$ has been found, the limit of $\tilde{r}_k^n$
is determined using a polynomial extrapolation (Neville's method) at $v^{-1}=0$.

\subsubsection{Degenerate eigenfunctions}\label{ssec:eig-lat}

This is the generic situation (generic superselection sectors) found in the
APS and sLQC prescriptions. In the basis construction, we follow (with minor
improvements) the procedure presented in Refs.~\cite{aps-det,aps-imp}. As already
mentioned, the eigenspaces are twofold degenerate, but in our analysis we concentrate ourselves on a
distinguished one-dimensional family of eigenstates (which can be next extended to the full basis via
orthogonal completion). The general eigenfunctions are solutions of a genuine second-order
difference equation, and hence require the specification of two pieces of initial data, e.g. the values at
two consecutive points of their support. The restriction by parity symmetry
does not impose any constraints on the data, since for generic sectors the image under parity reflection
of a lattice $\lat_{\varepsilon}$ is a different lattice $\lat_{4-\varepsilon}$. This
implies in particular that \emph{any} eigenfunction supported on $\lat_{\varepsilon}$
can be extended in a straightforward way to
$\lat_{\varepsilon}\cup\lat_{4-\varepsilon}$ by (anti)symmetry.

Taking into account all this, we construct the (distinguished half) basis
as follows:
\begin{enumerate}[(i)]
  \item \label{it:gen-bas} first, we build on $\lat_{\varepsilon}$ a pair of auxiliary eigenfunctions
    $\psi_k^\pm(v)$  [again solutions to Eq. \eqref{eq:e-def}] which
    converge to the WDW basis elements $\ub{e}_{-|k|}$ in the limits $v\to\pm\infty$, respectively;
  \item then, after a suitable rotation of their phases in a process detailed below, we add
  these functions and (delta-)normalize the outcome.
\end{enumerate}

In the first step of these computations, we choose the domain $I=\lat_{\varepsilon}\cap[-v_M,v_M]$,
where $v_M$ is selected as in Eq. \eqref{eq:vM}. The initial data for
$\psi^{\pm}$ are given at the lowest (for $-$) or greatest (for $+$) pair of points in $I$,
and are set \emph{equal} to the values of $\ub{e}_{-|k|}(v)$ at those
points. While this construction is not an exact implementation of \eqref{it:gen-bas} above, in practice
it approximates it with sufficient precision, owing to the quick convergence of the LQC
eigenfunctions to their WDW limits.

Once the auxiliary eigenfunctions are evaluated, we determine their WDW limit at
the opposite orientation side $v\to\mp\infty$. Since the initial data are complex,
this limit does not generally correspond to WDW standing waves, and takes the more
general form:
\begin{equation}
  \psi^\pm_k(v) = a_\varepsilon^\pm e^{i\alpha_\varepsilon^\pm}\underline{e}_{k}(v)
  + b_\varepsilon^\pm e^{i\beta_\varepsilon^\pm}\underline{e}_{-k}(v),
\end{equation}
where $a_\varepsilon^\pm,b_\varepsilon^\pm\in\mathbb{R}^+$, whereas
$\alpha_\varepsilon^\pm,\beta_\varepsilon^\pm\in [0,2\pi)$. They are all functions
of $k$, but we will ignore this in the notation so that it does not get too complicated.
The numerical analysis shows that the amplitude coefficients $a_\varepsilon^\pm$
and $b_\varepsilon^\pm$ grow (approximately) in an exponential way with $k$.
And, on the other hand, the selfadjointness of $\hat\Theta$ implies that
$|a^{\pm}_{\varepsilon}|^2-|b^{\pm}_{\varepsilon}|^2 = 1$ \cite{aps-det,aps-imp}.

To evaluate these coefficients, we split $\psi^{\pm}$ into real and imaginary
parts, denoted from now on by the symbols $\Re$ and $\Im$, respectively.
Since each of them separately converges to a standing wave, we can then directly apply
the technique used in the nondegenerate situation, presented in Sec.~\ref{ssec:eig-semi}.
The only complication with respect to that case is that, in addition to the norm factors
[like $\tilde{r}^{\varepsilon}_k$ in Eq. \eqref{eq:amp-phase-r}], we also need to find
the phase shifts $\phi^{\varepsilon}_k$. We do so by constructing the sequences
$\{(v_n^{-1},\phi^n_k)\}$ --analogs of $\{(v_n^{-1},\tilde{r}^n_k)\}$-- and by finding their
limit when $v_n^{-1}\to 0$. As before, the values of $\phi^n_k$ are determined for large $k$
by the positions of extrema, while for low $k$ they are evaluated algebraically.

Once we know the limiting coefficients of the four components $\Re[\psi^\pm]$ and
$\Im[\psi^{\pm}]$, the coefficients $a_\varepsilon^\pm$, $b_\varepsilon^\pm$,
$\alpha_\varepsilon^\pm$, and $\beta_\varepsilon^\pm$ can be easily calculated in terms of them. The determined data are then used to construct the desired linear combination of the two components
$\psi^{\pm}$. This involves two aspects: $(i)$ normalization, and $(ii)$ rotation.

Concerning the normalization, we rescale the function using the fact that
$4\|\psi_k^\pm\|^2 =|a_\varepsilon^\pm|^2 +|b_\varepsilon^\pm|^2+1$. This
ensures that each of the two considered components contributes with the same weight to the final basis element.
Thus, the final result will have a very similar behavior to that of the asymptotically
standing waves of Sec.~\ref{ssec:eig-semi}.

We then rotate $\psi^{\pm}$ to compensate for the overall phase
\begin{equation} \label{phas-rot}
  \chi_\varepsilon^\pm = -\frac{1}{2}(\alpha_\varepsilon^\pm +\beta_\varepsilon^\pm) .
\end{equation}
This step, new with respect to the procedure specified in Refs. \cite{aps-det,aps-imp},
is convenient to improve the semiclassicality properties of the physical states
constructed with our techniques from
the final basis elements.

As a result, we obtain the new two components $\tilde{\psi}_k^\pm(v) =
e^{i\chi_\varepsilon^\pm}\|\psi_k^\pm\|^{-1}\psi_k^\pm (v)$. One can see that,
in the corresponding limits $v\to\mp\infty$, they
behave as in Eq.~\eqref{wdw-limit} up \
to corrections of order $(a_\varepsilon^+)^{-1}$, which is a sufficiently good
approximation for $k \gg 1$.

Finally, these components are added and their sum is normalized:
\begin{equation} \label{latt-eigen}
  \tilde{e}^{\varepsilon}_{k}(v)
  = \frac{1}{\sqrt{2}}\frac{\tilde{\psi}^+_k(v)+\tilde{\psi}^-_k(v)}{\sqrt{1+\Re[z_k]}} ,
\end{equation}
where
\begin{equation} \label{in-cont}
  z_k = \frac{a^-_\varepsilon e^{i(\phi^-_\varepsilon-\chi^+_\varepsilon)}
      +a^+_\varepsilon e^{-i(\phi^+_\varepsilon-\chi^-_\varepsilon)}}
    {\|\psi_k^+\|\,\|\psi_k^-\|} .
\end{equation}
This last quantity comes from the scalar product between $\tilde\psi_k^+$ and $\tilde\psi_k^-$,
and in the regime $k \gg 1$ is of the order of $(a_\varepsilon^+)^{-1}$. As a consequence, it can be neglected
for physically interesting states (large $k$). In such case, one again recovers
the behavior of $\tilde{e}^{\varepsilon}_{k}$ shown in Eq.~\eqref{wdw-limit}.

The final step in the basis construction is the symmetrization to get the generalized eigenfunction $(e^{\varepsilon}_{k}|$
supported on $\mathcal{L}_\varepsilon\cup \mathcal{L}_{4-\varepsilon}$:
\begin{equation} \label{sim-eigen}
  (e^{\varepsilon}_{k}|v\rangle = \frac{1}{\sqrt{2}} \left[ (\tilde{e}^{\varepsilon}_{k}|v\rangle
  + (\tilde{e}^{\varepsilon}_{k}|-v\rangle \right] .
\end{equation}

\subsection{Physical states: construction and analysis}\label{sec:num-states}

In the numerical analysis that was carried out in Refs.~\cite{aps-det,aps-imp}, the physical
states consisted in Gaussian distributions
\begin{equation}\label{gauss-state}
  \tilde{\Psi}(k) = \Psi_G(k) = \frac{1}{(2\pi)^{1/4}\sqrt{\sigma}}e^{-(k-k_0)^2/(4\sigma^2)},\quad k\in \mathbb{R}.
\end{equation}
The parameters $k_0$ and $\sigma$ are related in a simple way with the expectation
value $\langle \hat p_\phi\rangle$ and the dispersion $\Delta \hat{p}_\phi$ of the momentum of the scalar field
\begin{equation}
  \langle\hat p_\phi\rangle=\sqrt{12\pi G}\, k_0, \qquad
  \frac{\Delta \hat{p}_\phi}{\langle\hat p_\phi\rangle} = \frac{\sigma}{k_0}.
\end{equation}
In principle, the support of the Gaussian spectral profile is the entire real
line, thus being directly applicable to the cases when the basis elements cover the
entire set $k\in\re$ (the degenerate situation described in Sec.~\ref{ssec:eig-lat}, after
including the orthogonal complement of the constructed half basis).
In the cases where the spectrum of $\hat{\Theta}$ is nondegenerate, the Gaussian
profiles suffer a modification (owing to the cutoff at $k=0$). Therefore, in such situations
the final shape resembles a true Gaussian only for profiles that are sharply peaked,
so that $k_0$ is large compared to $\sigma$.

Since we are interested in the study of more general physical states, for which the different prescriptions
may in principle lead to different quantum predictions, we introduce more convenient profiles, applicable without
modifications to both the degenerate and the nondegenerate cases. Specifically, we consider logarithmic normal
distributions of the type:
\begin{equation} \label{loggauss-state}
  \Psi_{L}(k)=\frac{1}{(2\pi)^{1/4}\sqrt{\sigma k}}e^{-[\ln(k/k_0)]^2/(4\sigma^2)},
\end{equation}
with $k$ running over the positive semiaxis.
The positive parameters $k_0$ and $\sigma$ are related now to $\expect{\hat{p}_{\phi}}$ and $\Delta \hat{p}_{\phi}$
as follows:
\begin{equation}
  \langle\hat p_\phi\rangle=\sqrt{12\pi G}k_0e^{\sigma^2/2}, \qquad
  \frac{\Delta \hat{p}_\phi}{\langle \hat p_\phi\rangle} = \sqrt{e^{\sigma^2}-1}.
\end{equation}
We will analyze this two-parameter family of states to investigate the discrepancies between
prescriptions in the regimes where $\langle \hat p_\phi\rangle$ and $\Delta \hat{p}_\phi$ are of
the same order.

The wave function $\Psi(v,\phi)$ corresponding to a given profile $\tilde{\Psi}(k)$
can be evaluated directly by performing the integral \eqref{eq:phi-states}. Obvious
technical limitations require us to, first, discretize the integral and, second,
restrict it to a compact domain $\mathbb{D}$ in $k$. For our purposes, it is
sufficient to choose $\mathbb{D}$ as
\begin{equation}
  \mathbb{D}=[k_0e^{-s\sigma},k_0e^{s\sigma}],\quad {\rm with} \quad s\in\natu^+.
\end{equation}
As far as $s>7$, one can check that the relative error in the integration owing to the neglected contribution of
$k\in\re\setminus\mathbb{D}$ is less than $10^{-12}$.
In our simulations, we have chosen $s=10$.

For the numerical integration of Eq. \eqref{eq:phi-states} in $\mathbb{D}$, we have used
\emph{Romberg's} method (see e.g. Ref. \cite{romb-refs}).
This method is particularly convenient if one wants to restrict the number of integrand
probing points in $\mathbb{D}$, which is the case here, as the evaluation of the basis
elements is the most numerically expensive step of the process. To control the
integration precision, we have demanded that the difference of the results between the consecutive
orders $l$ and $l+1$ of the polynomial extrapolation (an internal component of the Romberg
method \cite{romb-refs}) satisfy
\begin{equation}
  \|\Psi^{(l+1)}_\phi-\Psi^{(l)}_\phi\|_{\phy} < \delta \|\Psi^{(l+1)}_\phi\|_{\phy},
\end{equation}
where the imposed error bound $\delta$ varies from $10^{-6}$ to $10^{-10}$, depending on
the simulation. To avoid an excessive cost of time in the integration, we have restricted the number of the
integrand probing points (forming the uniform lattice in $\mathbb{D}$)
to $2^{12}+1$.

\begin{figure*}[tbh!]
  \subfigure[]{
    \includegraphics[scale=0.65]{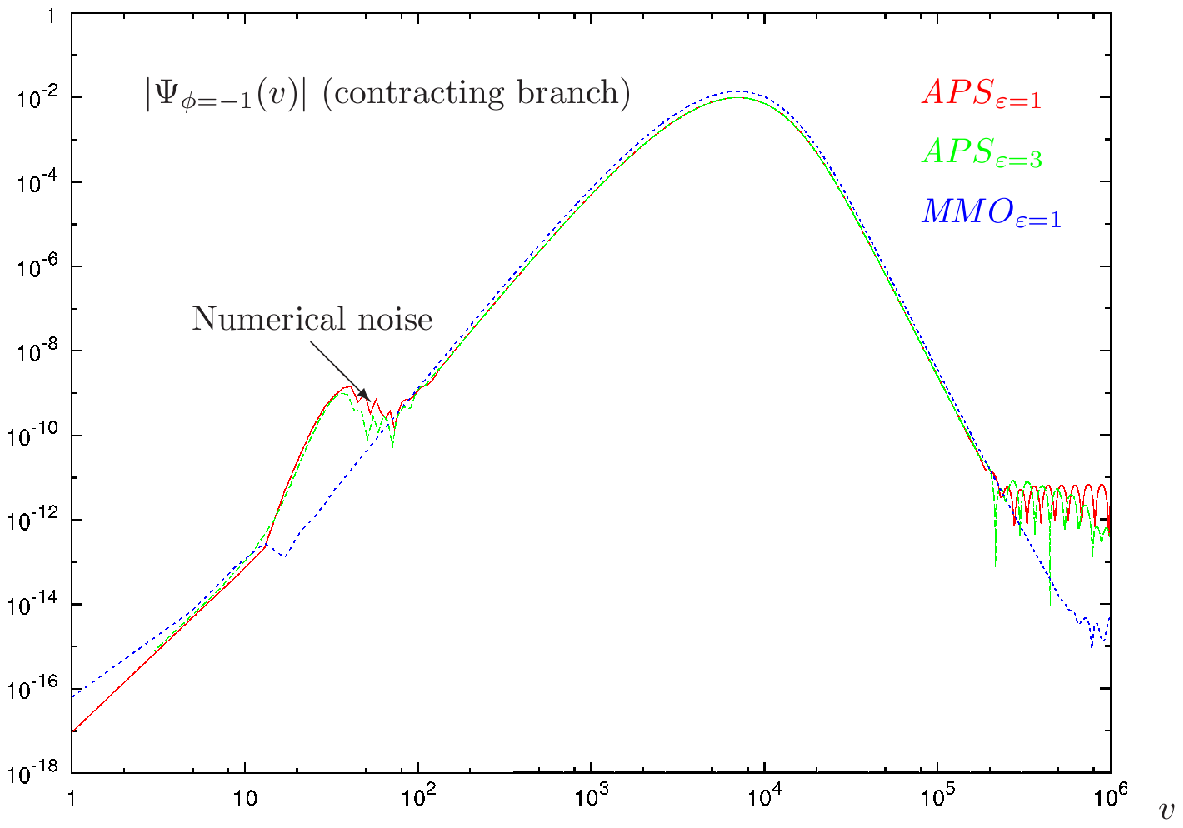}
    \label{fig:prof-class}
  }
  \subfigure[]{
    \includegraphics[scale=0.65]{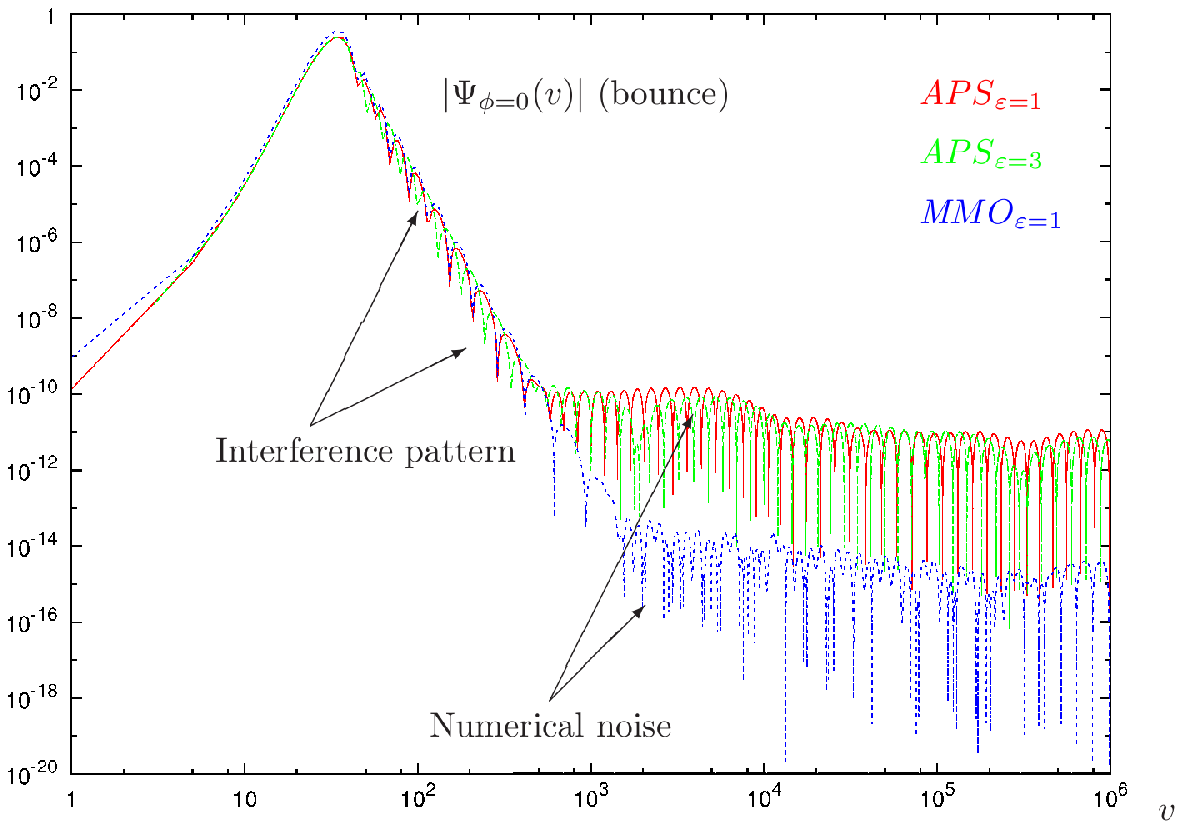}
    \label{fig:prof-bounce}
  }
  \caption{Amplitude $|\Psi_{\phi}(v)|$ of the physical wave function
    corresponding to a state with a logarithmic normal distribution. The amplitude for different
    prescriptions is compared both away from the bounce $(a)$ and at the bounce $(b)$. The parameters
    of the profile $\tilde{\Psi}$ of this state are fixed by the conditions
    $\langle \hat p_\phi\rangle=100\hbar$ and $\Delta \hat{p}_\phi/\langle \hat p_\phi\rangle=0.1$.
    Away from the bounce, the amplitudes are indistinguishable up to numerical noise, whereas
    at the bounce one can observe differences (phase shift) in the interference pattern.
    The noise level clearly depends on the used techniques, something which depends in turn on the degeneracy
    of the spectrum of $\hat{\Theta}$.
  }
  \label{fig:prof}
\end{figure*}

\subsection{Observables}\label{sec:num-obs}

We can now proceed to calculate the action of
observables on the states represented by the wave functions constructed in the previous subsection.
In particular, we can evaluate and compare
the expectation values and dispersions of those observables. We consider two types of
observables: those introduced in Sec.~\ref{ssec:obs-def}, which encode standard properties of interest in
cosmology  --namely, $\ln|\hat{v}|_{\phi}$, $\hat{H}_{\phi}$, and $\hat{\rho}_{\phi}$--,
and the observables defined in Sec.~\ref{sec:presc-diff}, which are specially suitable to detect the
differences between the studied prescriptions.

The dynamical information is extracted by means of the Schr\"odinger
picture, where the evolution of a state is seen as a mapping between
initial data (on a constant $\phi$ slice) via the unitary transformation \eqref{eq:evol-map-unit}. In this
picture, the action of a physical observable is obtained from that of its
kinematical precursor (see Sec.~\ref{ssec:obs-def}) on the appropriate initial
data slice.

This fact has been applied in previous numerical studies of
LQC, starting with Ref. \cite{aps-imp}, in order to extract dynamical data.
In our case, in the $v$ representation, all the interesting kinematical operators (precursors)
are either multiplication operators or combinations of multiplications
and shifts. This simple form allows us to evaluate the results of their action
straightforwardly by making use of the map $\Psi_{\phi}(v)\mapsto \ln|v|\Psi_{\phi}(v)$, Eq. \eqref{eq:rho-def},
Eq. \eqref{eq:H-def}, and the right formula in Eq. \eqref{eq:DeltaTheta}.

The expectation values are then evaluated via the kinematical inner product
on $\Hil_{\gr}$,
\begin{equation}\label{eq:num-IP}
  \langle\Psi_{\phi}|\Phi_{\phi}\rangle
  = \sum_{v\in\lat_{\epsilon}\cap J} \bar{\Psi}_{\phi}(v)\Phi_{\phi}(v) \ ,
\end{equation}
where, for technical reasons, the summation is restricted to the compact region
$J=[-v_{m'},v_{m'}]$ (in the degenerate case) or $J=[0,v_{m'}]$ (in the nondegenerate
case). In our simulations the bound $v_{m'}$ has been selected to lay always in
the interval $[10^4,4\cdot 10^6]$, its specific value varying for different simulations.
This choice ensures that the error caused by the restriction to a compact domain $J$
has a subleading contribution compared with other numerical
errors that arise in the evaluation process.

To isolate the numerical noise (see Fig.~\ref{fig:prof}) generated
by the errors introduced in the evaluation of the basis and in the integration of
the wave function $\Psi$ on each slice, the values of $\Psi_{\phi}(v)$ entering Eq.~\eqref{eq:num-IP}
have been modified
by a filter, namely, whenever $|\Psi_{\phi}(v)|<\boldsymbol{\alpha} \sup |\Psi_{\phi}|$
the value $\Psi_{\phi}(v)$ has been set equal to zero. This prevents this type of noise from affecting the
computation of the expectation values. In our simulations, the value of the relative bound
$\boldsymbol{\alpha}$ has been selected to vary between $10^{-8}$ and $10^{-6}$.

The dispersions have been calculated using the standard formula
\begin{equation}
  \expect{\Delta\hat{O}}^2 = \expect{\hat{O}^2} - \expect{\hat{O}}^2
\end{equation}
for each observable $\hat{O}$.

Finally, since we work in the symmetric sector, we note that we can restrict all our considerations to
half of the support of the wave function; in particular, when limited to a compact region, we can restrict
ourselves to a domain like $J$, as specified below Eq. \eqref{eq:num-IP}.\\

\subsection{Efficiency and precision}\label{sec:effic}

Let us now investigate the properties of the numerical techniques discussed in the
previous subsections from the perspective of the computational precision and efficiency.
The numerical computations necessary to get the final results consist of several steps:
$(i)$ evaluation of the basis elements $(e_k^\varepsilon|$ (discussed in detail in
Sec.~\ref{sec:num-basis}), $(ii)$ integration of the spectral profile to determine the
wave function for each constant value of $\phi$ (Sec.~\ref{sec:num-states}), and $(iii)$
evaluation of the action of the observables and computation of their expectation values
and dispersions (Sec.~\ref{sec:num-obs}). Each of these steps introduces its own source of
numerical error and presents a different efficiency.

We start with the first of these steps: the evaluation of the basis elements. The
comparisons during the simulations have shown that this step is responsible for most of
the computational cost, and therefore it is the most critical part from the viewpoint of
the efficiency. As discussed in Sec.~\ref{sec:num-basis}, the actual algorithms and cost
depend on the degeneracy of the basis, and hence vary significantly with the considered
superselection sector and quantization prescription.

In the nondegenerate case (Sec.~\ref{ssec:eig-semi}), the calculation involves two steps:
the determination of the non-normalized eigenfunctions $\psi_k$ and the computation of
their norm by finding their WDW limit. The calculation precision depends on the size of
the evaluation domain chosen for the eigenfunction and on the wave number $k$ [see Eq.
\eqref{eq:phi-states}]. In particular, we observe that two effects compete: since the
eigenfunctions are evaluated via iterative methods, the evaluation precision decreases
with the size of the domain, whereas the precision in determining the WDW limit increases
with it. In that respect, the choice of the domain size given by Eq. \eqref{eq:vM}
provides a fairly acceptable balance between these two sources of error. It is also worth
recalling that, with our conventions [$e_k^{\varepsilon}(\varepsilon)>0$], there are no
ambiguities in the freedom of choice for the global phase of the eigenfunctions.

The degenerate case, as we have seen in Sec.~\ref{ssec:eig-lat}, is considerably more
complicated. First, the procedure applied in the nondegenerate case becomes just the first
step of the evaluation. Even this stage introduces now a higher numerical error, because
the domain of calculation of $\psi_k$ is now \emph{twice} larger, and hence the evaluation
of the eigenfunctions requires twice more iterative steps. Apart from that, we observe a
significant cost increase since we have to evaluate the \emph{pair} of eigenfunctions
$\psi^{\pm}_k$ and, besides, both $\psi^{\pm}_k$ are now \emph{complex} instead of real.
In total, the three commented facts amount to an increase of {\bf 8 times} in the
computational cost.

Furthermore, the next step --taking the appropriate linear combination of $\psi^{\pm}$ to
form the final basis functions-- has its own cost (which is linear in the domain size).
Apart from that, the rotational symmetry of the components $\psi^{\pm}$ is broken, in the
sense that, in order to construct the appropriate basis vectors, we need to compensate for
the overall phase of the WDW limits of those components \eqref{phas-rot}. This step
introduces extra complications, since the phase itself can be determined only modulo
$\pi$. The correct identification of this phase, crucial for the subsequent construction
of the relevant physical states, is nontrivial, and in fact one can check that this
phase is proportional to $k\ln|k|$ at its leading order. As a consequence, this step in
the determination of $e_k^{\varepsilon}(v)$ introduces an additional numerical error.

In the integration of the wave function profiles [step $(ii)$ above], the use of the
high-order Romberg's method allows us to restrict the number of evaluated basis elements
to a manageable amount, while keeping sufficiently high numerical precision. The selection
of this method and of a proper compact integration domain makes also possible that both
the integration error and the error caused by the restriction of the domain can be limited
to a level where they do not exceed the error generated in our previous step $(i)$ of the
numerical computation. The differences between the degenerate and nondegenerate cases do
not require a different treatment. However, in practice, the degenerate situation
turns out to be approximately {\bf $3$ times} more expensive numerically owing to
two reasons: $(a)$ because the
eigenfunctions $e_k^{\varepsilon}(v)$ are complex in that case, and $(b)$ because the
wave function has to be calculated for both $v>0$ and $v<0$.

The effect and dependence of the overall numerical error introduced in the previous steps
$(i)$ and $(ii)$ is shown in Fig.~\ref{fig:prof}. For the states analyzed in this article,
the error stays at the level of $10^{-12}$ in the nondegenerate case. The additional
complications characteristic of the degenerate case cause the error to grow in those cases
by $2$ or $3$ orders of magnitude. Nonetheless, all the wave function profiles can be
integrated with a final relative error which does not exceed $10^{-8}$.

The final step $(iii)$ in the numerical computation involves algorithms which are common
for both the degenerate and the nondegenerate cases. However, in the degenerate case, a
higher level of numerical noise is visible in Fig.~\ref{fig:prof}. This has forced us to
conveniently increase in this case the value of the relative bound $\boldsymbol{\alpha}$
in the discrimination filter (see Sec.~\ref{sec:num-obs}). In turn, this happens to
increase the error in the evaluation of the expectation values and dispersions by the same
order of magnitude (i.e., it increases from approximately $10^{-12}$ to
$10^{-9}$--$10^{-8}$).

\section{Results and discussion}\label{sec:results}

\begin{figure*}
  \subfigure[]{
    \includegraphics[scale=0.65]{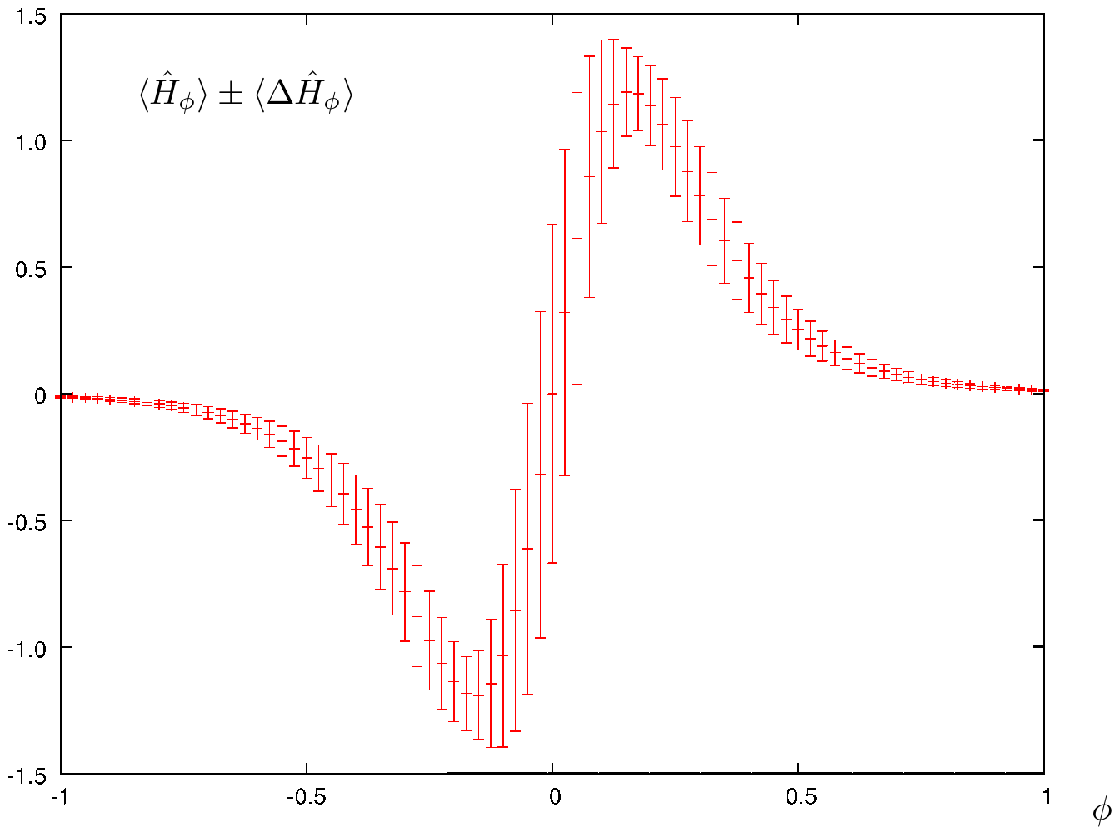}
    \label{fig:H-traj}
  }
  \subfigure[]{
    \includegraphics[scale=0.65]{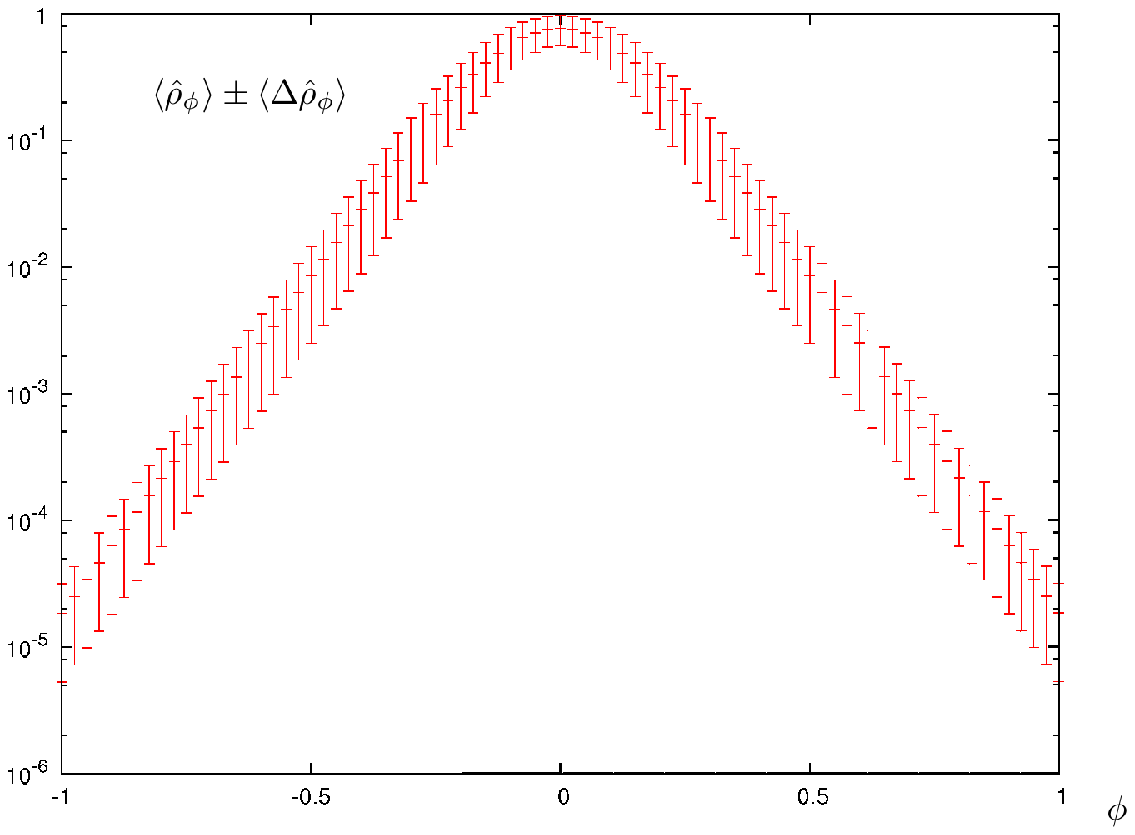}
    \label{fig:rho-traj}
  }
  \caption{Dynamical trajectories of $\hat{H}_\phi$ $(a)$ and $\hat{\rho}_\phi$ $(b)$, given
    by the expectation values of these observables on the state of
    Fig.~\ref{fig:prof} with $\varepsilon=1$ in the APS prescription.
  }
  \label{fig:Hrho-traj}
\end{figure*}

\begin{figure*}
  \subfigure[]{
    \includegraphics[scale=0.65]{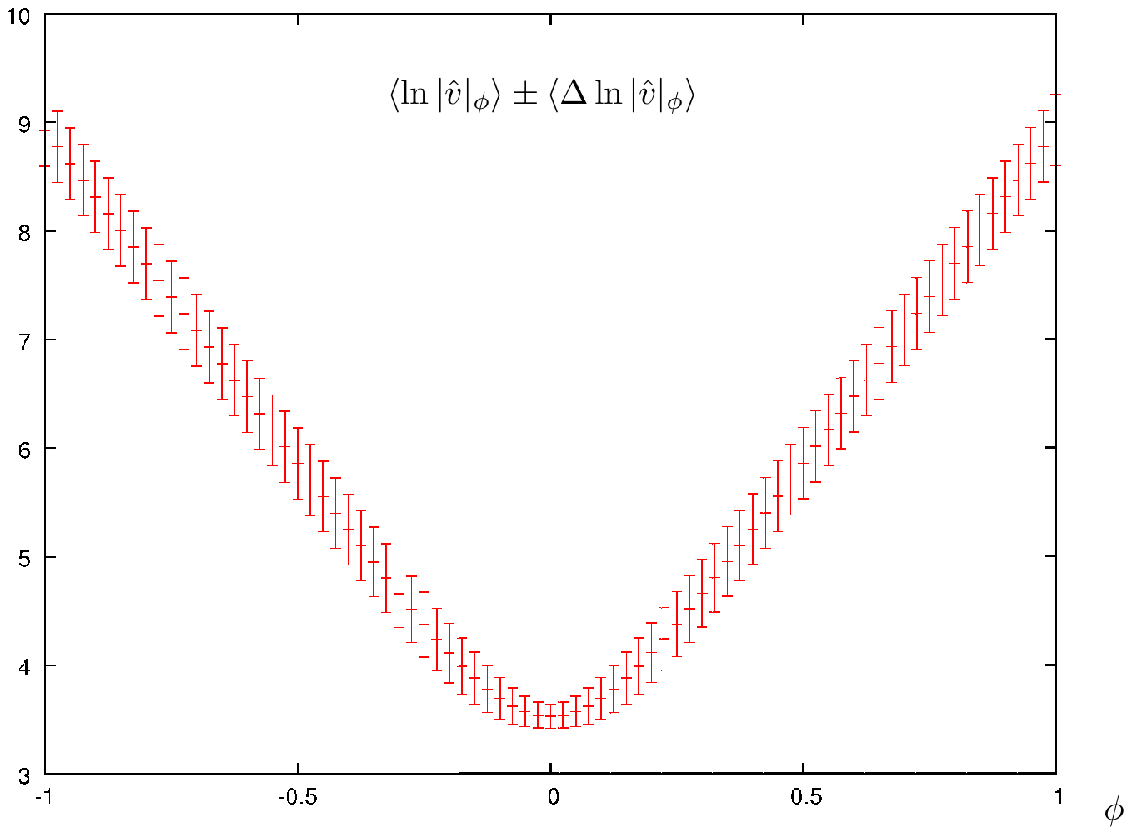}
    \label{fig:logv-traj}
  }
  \subfigure[]{
    \includegraphics[scale=0.65]{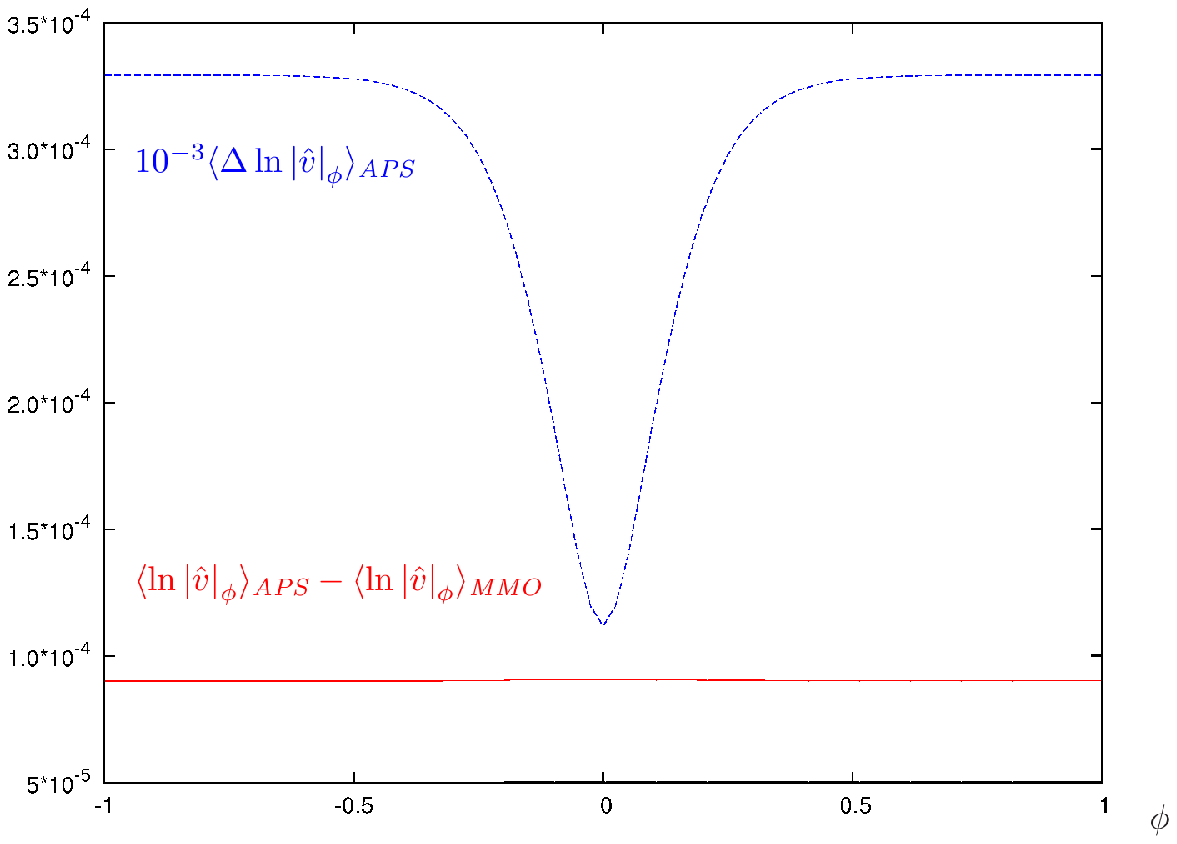}
    \label{fig:logv-diff}
  }
  \caption{$(a)$ Quantum trajectory of $\ln|\hat{v}|_\phi$ for the same state and the same prescription as in Fig.~\ref{fig:Hrho-traj}.
    $(b)$ Uncertainty in $\ln|\hat{v}|_\phi$ for the same state and prescription, compared with
    the difference between the corresponding expectation values calculated in the APS
    and MMO prescriptions.
  }
  \label{fig:logv}
\end{figure*}

\begin{figure*}
  \subfigure[]{
    \includegraphics[scale=0.65]{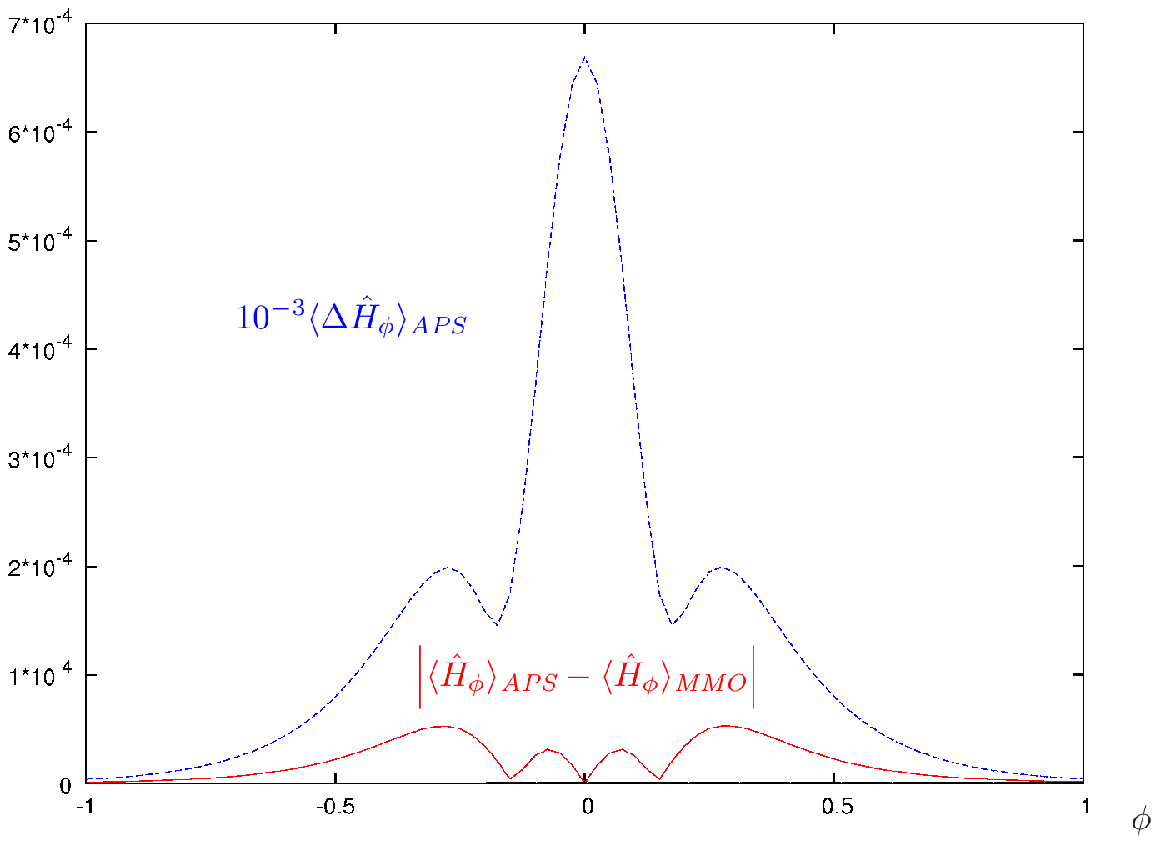}
    \label{fig:Hdiff-abs}
  }
  \subfigure[]{
    \includegraphics[scale=0.65]{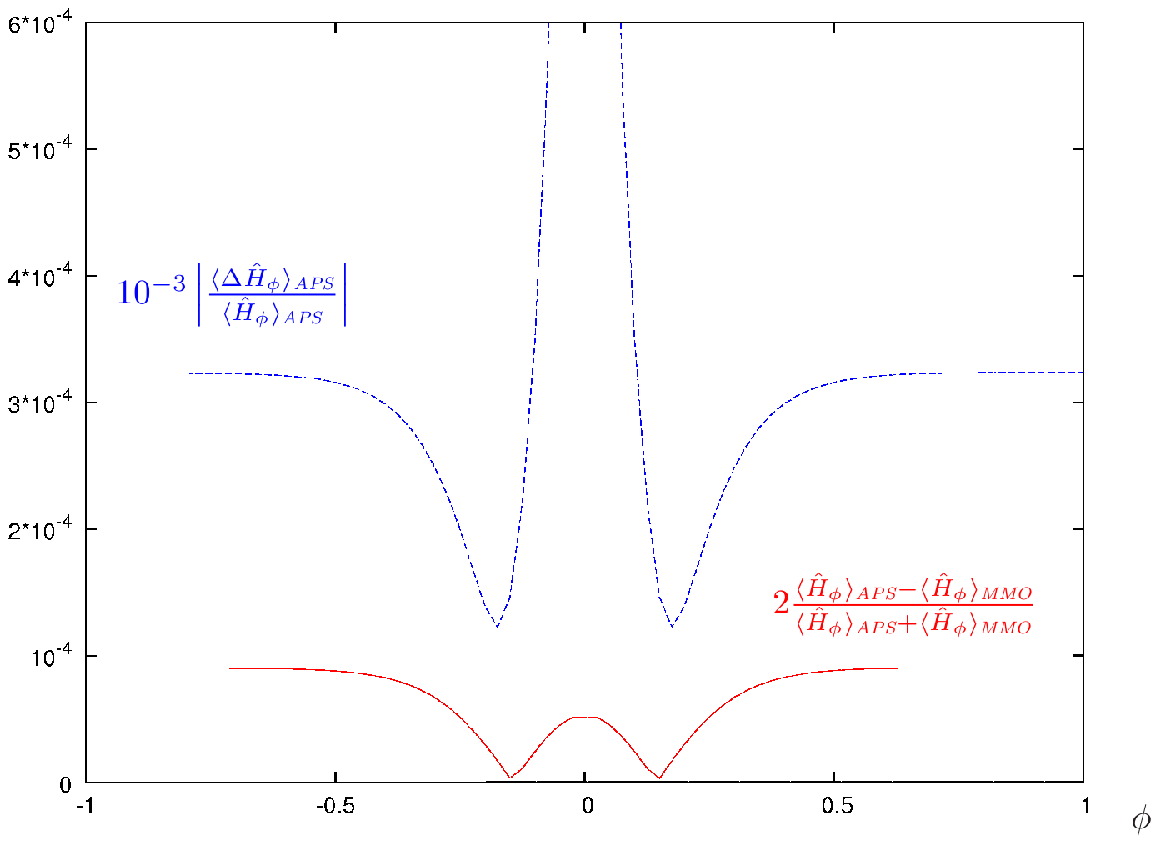}
    \label{fig:Hdiff-rel}
  }
  \caption{Absolute dispersions $(a)$ and relative dispersions $(b)$ of $\hat{H}_{\phi}$ for
    the considered state of Fig.~\ref{fig:prof}, compared with the corresponding difference between the expectation
    values of $\hat{H}_{\phi}$ in the APS and MMO prescriptions. For both relative values, one can observe
    a peak at the bounce owing to the vanishing of
    $\langle\hat{H}_\phi\rangle$; however, the peak in the differences (red curve) is so sharp
    that it is placed between probing points.
  }
  \label{fig:Hdiffs}
\end{figure*}

\begin{figure*}
  \subfigure[]{
    \includegraphics[scale=0.65]{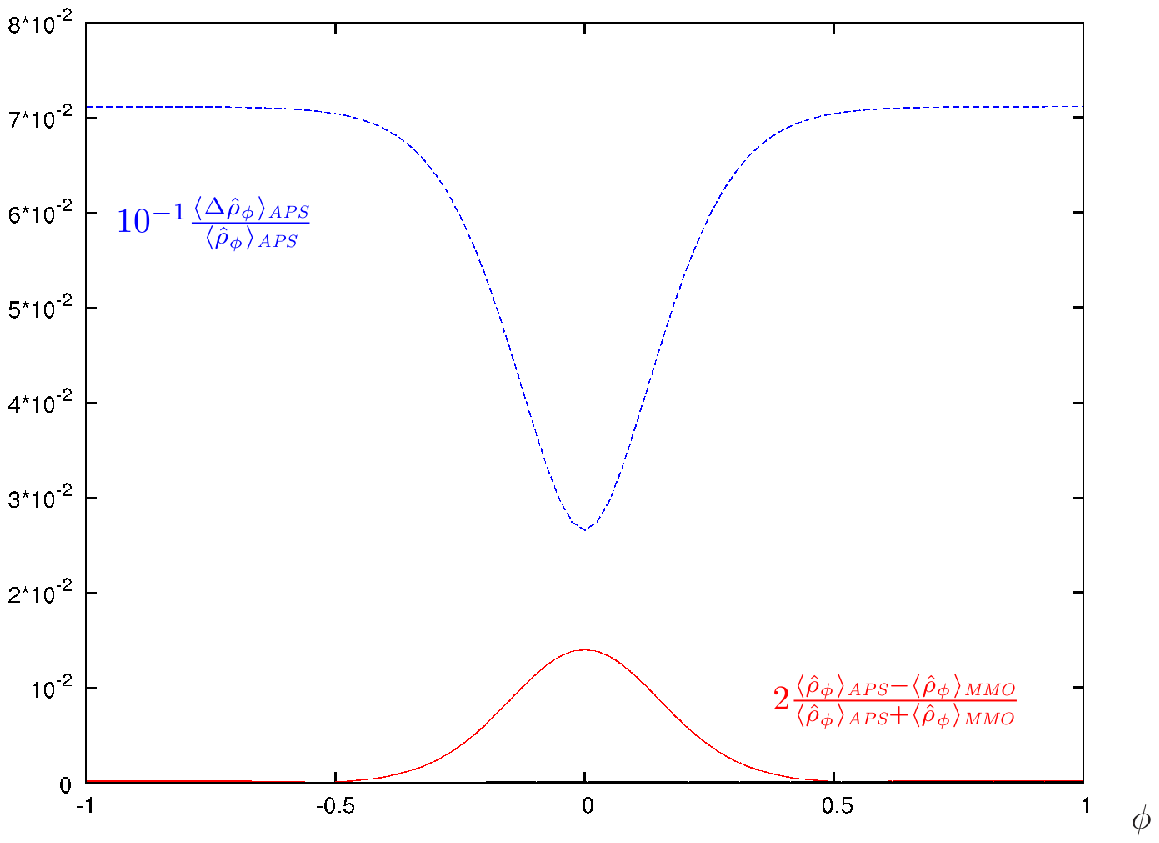}
    \label{fig:rho-diff}
  }
  \subfigure[]{
    \includegraphics[scale=0.65]{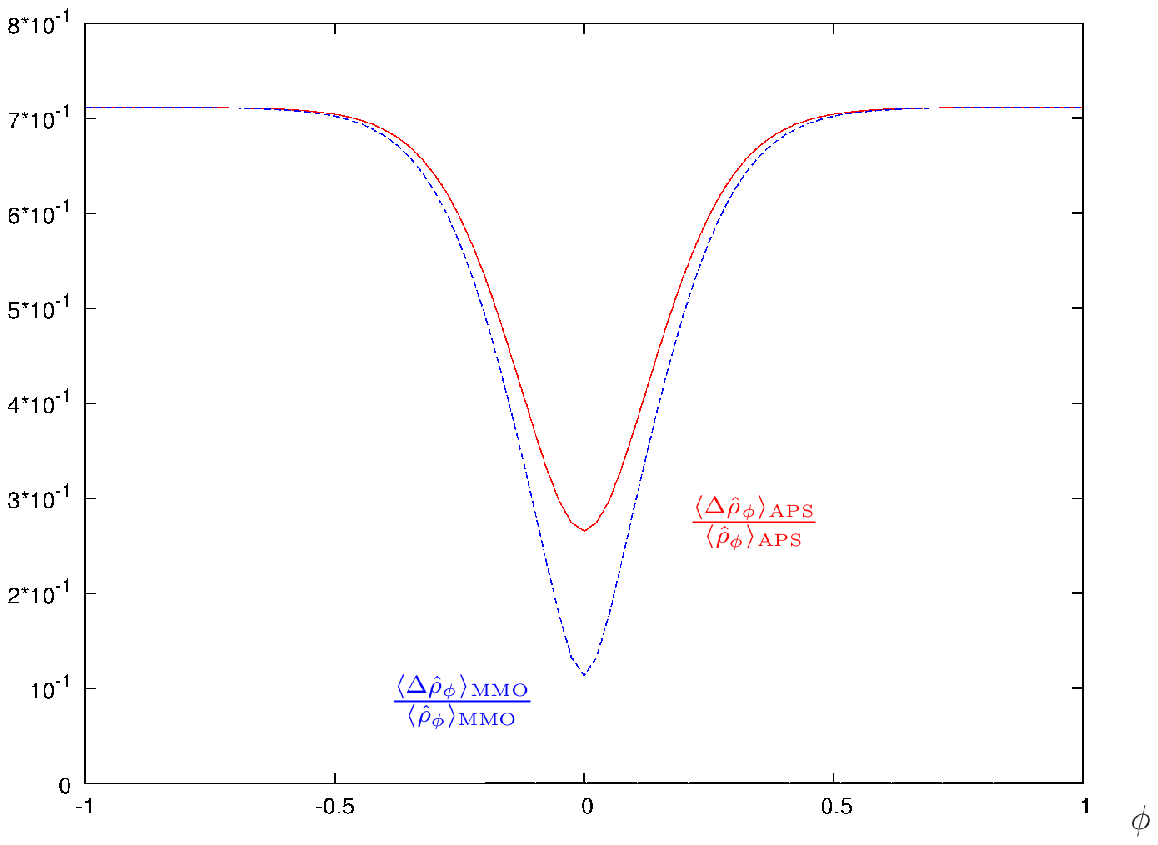}
    \label{fig:rho-gendiff}
  }
  \caption{$(a)$ Relative dispersion of $\hat{\rho}_{\phi}$ for the same state as in the previous figures,
    compared with the relative difference between the corresponding expectation values in the APS
    and MMO prescriptions.
    $(b)$ Comparison between the relative dispersion of $\hat{\rho}_{\phi}$ in the APS and MMO prescriptions, for a
    ``generic'' superselection sector.
    Better coherence properties are observed for the state constructed in the MMO prescription.
  }
  \label{fig:rhodiffs}
\end{figure*}

\begin{figure*}[tbh!]
  \subfigure[]{
    \includegraphics[scale=0.65]{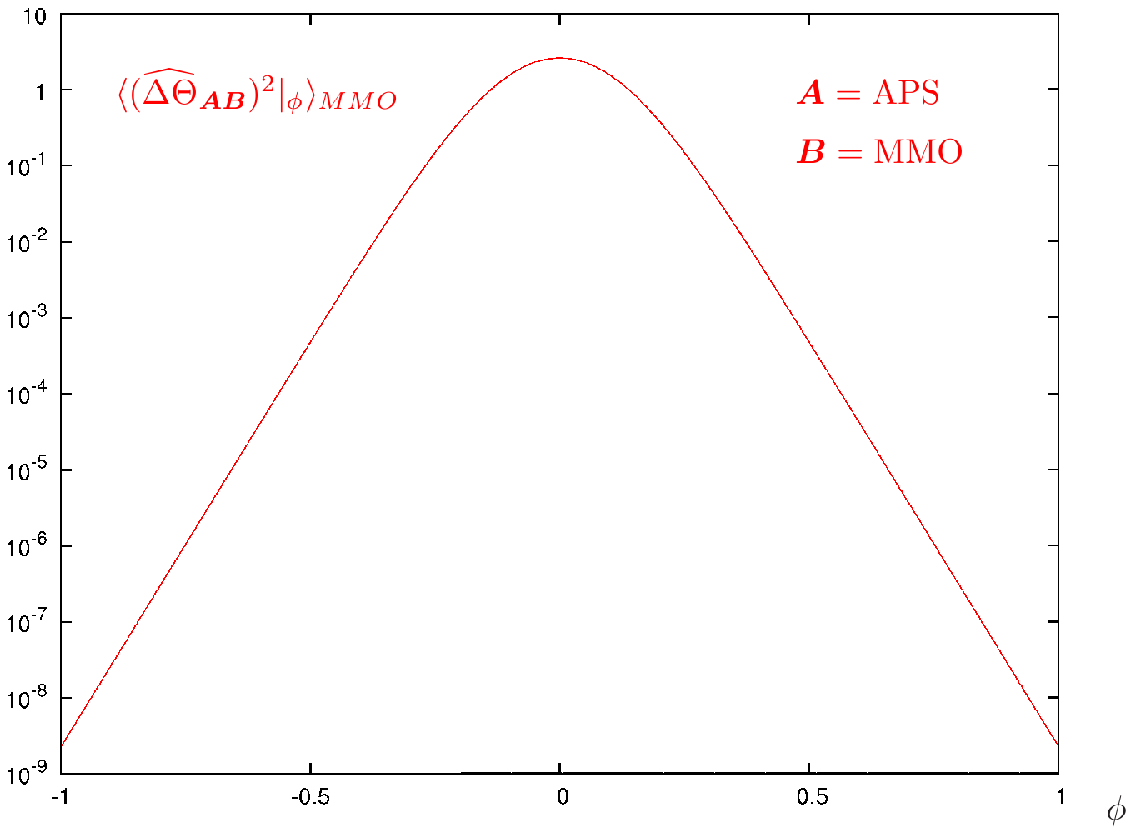}
    \label{fig:thetadiff-APS}
  }
  \subfigure[]{
    \includegraphics[scale=0.65]{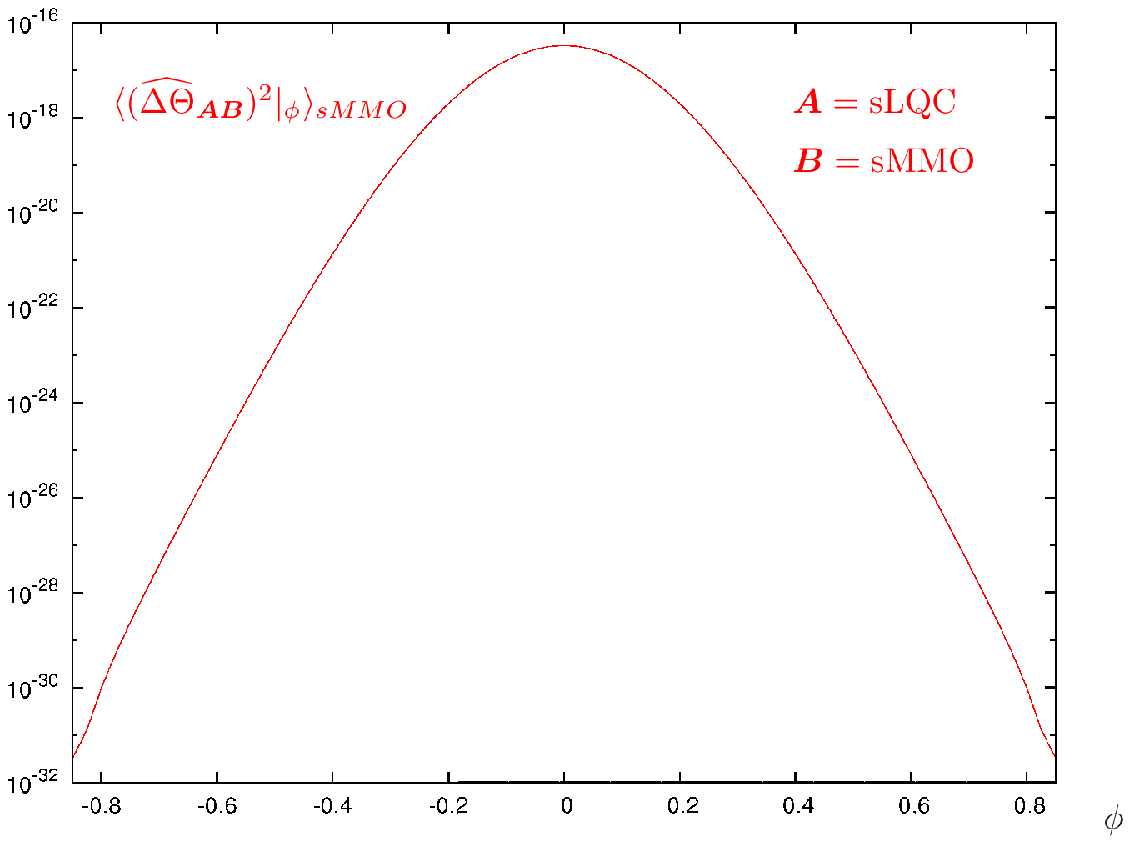}
    \label{fig:thetadiff-MMO}
  }
  \caption{Expectation values of the observable
    $(\widehat{\Delta\Theta}_{\boldsymbol{AB}})^2|_{\phi}$ for $\boldsymbol{A}=\APS$ and
    $\boldsymbol{B}=\MMO$ $(a)$; as well as for $\boldsymbol{A}=\sLQC$ and $\boldsymbol{B}=\sMMO$
    $(b)$. In both cases, $\varepsilon=1$ and the observable is evaluated on the state with a logarithmic normal distribution
    whose parameters are $\langle \hat p_\phi\rangle=100\hbar$ and
    $\Delta \hat{p}_\phi/\langle \hat p_\phi\rangle=0.1$. The state is built in the MMO prescription in case $(a)$, and in
    the sMMO prescription in case $(b)$. The difference reaches a maximum at the
    bounce and decays exponentially away from it. The difference between the sLQC and the sMMO prescriptions
    is many orders of magnitude smaller than the difference between any other pair of prescriptions.
  }
  \label{fig:thetadiff}
\end{figure*}

We have applied the methods explained in the previous section to the numerical analysis of
a population of states that are given by a normal logarithmic distribution of the form
\eqref{gauss-state}, with the value of $\langle\hat{p}_{\phi}\rangle$ ranging from
$30\hbar$ to $500\hbar$ and the relative dispersion
$\langle\Delta\hat{p}_{\phi}\rangle/\langle\hat{p}_{\phi}\rangle$ from $0.05$ to $0.25$.
The analysis of these states has been carried out in the four prescriptions discussed in
this article. We have analyzed $2^5$ different values of $\varepsilon$, labeling distinct
superselection sectors. The results are displayed in
Figs.~\ref{fig:prof}-\ref{fig:thetadiff}. At various levels of comparison, we can
distinguish the following aspects.

First, a preliminary comparison can be performed at the level of the wave function itself.
Namely, one can compare the probability amplitude --the value $|\Psi(v,\phi)|$ scaled by
the square root of the inner product measure on $\Hil_{\gr}$-- of the wave function which
represents the same state [i.e., with the same spectral profile $\tilde{\Psi}(k)$] in the
different prescriptions. Away from the bounce [see Fig.~\ref{fig:prof-class}], one does
not observe any significant difference. However, at the bounce
[Fig.~\ref{fig:prof-bounce}] the situation becomes slightly more complicated. The general
shape of the wave function (position of the peak, general behavior of the function slopes)
still does not show any clear distinction; nevertheless, one observes a phenomenon that
actually reveals the existence of fine differences. In fact, the interaction of the
expanding and contracting branches creates an interference pattern, which can be seen on
the downward slope of the function in Fig.~\ref{fig:prof-bounce}. For a chosen
superselection sector and wave functions representing the same state, the pattern shows a
dependence on the prescription: for different prescriptions the minima and maxima of the
interference are displaced. Nonetheless, the specific shift of the various extrema depends
not only on the prescription, but also on the shape of the state (spectral profile), as
well as on the superselection sector to which it belongs. As a consequence, it cannot be
used in a systematic obvious way to identify the prescription employed, regardless of the
state under consideration.

Another, more physically relevant aspect for comparison has to do with the use of the
cosmological observables $\ln|\hat{v}|_{\phi}$, $\hat{H}_{\phi}$, and $\hat{\rho}_{\phi}$.
The results are presented in Figs.~\ref{fig:Hrho-traj}-\ref{fig:rhodiffs}. Analyzing the
same state in different prescriptions we have found detectable differences between the
expectation values of all the three observables [see Figs.~\ref{fig:logv-diff},
\ref{fig:Hdiffs}, \ref{fig:rho-diff}]. These differences are most prominent at the bounce
and decay quickly as the wave packet enters the low energy density regime. For the states
investigated here, for which $\langle\hat{p}_{\phi}\rangle \gtrsim 30\hbar$ and
$\langle\Delta\hat{p}_{\phi}\rangle \lesssim 0.25 \langle\hat{p}_{\phi}\rangle$, the
differences are nevertheless several orders of magnitude smaller than the dispersions of
the corresponding observables through all the evolution. Those differences depend on the
degeneration of the spectrum of $\hat{\Theta}$, apart from their natural dependence on the
observables and the particular state under consideration. The situation where the highest
differences have been observed occurs when one compares the expectation values of the
energy density operator on highly dispersed states with low momentum
$\langle\hat{p}_\phi\rangle$. Then, the computed differences lay only one or two orders of
magnitude below the dispersions during the whole evolution. In the rest of situations
considered here, the differences are even smaller when compared to the dispersions.

Among the results presented above, the dispersion of the energy density
$\hat{\rho}_{\phi}$ deserves special attention. For all the prescriptions, the essential
spectrum of this operator is the interval $[0,\rho_c]$ where $\rho_c\approx
0.81\rho_{\Pl}$ is the so called \emph{critical energy density}. Depending on the
prescription, the spectrum may also have a discrete part with eigenvalues exceeding
$\rho_c$, but these play no role in the states that represent the cosmological solutions
\cite{klp-aspects}. This fact is reflected in the behavior of
$\langle\Delta\hat{\rho}_{\phi}\rangle$. Namely, for the states analyzed in this paper and
for the nondegenerate cases (as defined in Sec.~\ref{sec:num-basis}), the expectation
value $\langle\hat{\rho}_{\phi}\rangle$ reaches the critical energy density $\rho_c$ at
the bounce, and its dispersion drops significantly there (in particular, it
\emph{vanishes} up to the numerical error for states peaked at large $p_{\phi}$).
In this sense, the states with the spectral profile \eqref{gauss-state} are coherent ones.
In the degenerate cases the situation is different: we observe that
$\langle\Delta\hat{\rho}_{\phi}\rangle$ decreases near the bounce, but it reaches a
positive minimum significantly larger (at least a few times) than in the nondegenerate
case. This property clearly distinguishes between the APS and sLQC prescriptions, on the
one hand, and the MMO and sMMO prescriptions, on the other hand, at least for the
superselection sectors with $\varepsilon\neq 0,2$. The observed difference might be
nonetheless related to the particular way of constructing the basis in the degenerate
cases \footnote{Recall that the freedom of choice for $e_k^{\varepsilon}(v)$ amounts to a
two-dimensional space.}.

The other physical aspect considered in our numerical analysis concerns the expectation
values of the observables $(\widehat{\Delta\Theta}_{\boldsymbol{AB}})^2$,
constructed specially to measure the discrepancies between the different prescriptions.
The results are presented in Fig.~\ref{fig:thetadiff}. As we can see, these expectation
values are nonvanishing. Thus the prescriptions are clearly \emph{distinct}, and the
differences between them are of course \emph{detectable}. As we could have guessed, the
expectation values (and therefore the physical differences between the prescriptions) are
largest near the bounce. Away from it, they decay exponentially. This behavior can in fact
be proven analytically for all physical states, for which
$\langle\Delta\hat{p}_{\phi}\rangle$ is finite, by employing methods similar to those
applied in Sec.~VB of Ref. \cite{kp-scatter}. Not surprisingly, the largest differences
are observed between prescriptions which lead to a different potential term in the
expansion \eqref{eq:theta-potential} (i.e., to different values of the constant $\alpha$).
An example of such situation is presented in Fig.~\ref{fig:thetadiff-APS}. In the case of
the sLQC and the sMMO prescriptions [Fig.~\ref{fig:thetadiff-MMO}], the difference is many
orders of magnitude smaller (in the presented case, more than 16 orders), because the
potential terms in these two prescriptions coincide and the only difference is a compact
operator, supported only on three lattice points near the classical singularity.

\section{Conclusions}\label{sec:concl}

In LQC, even in simplest models, the standard ambiguities of the canonical quantization
affecting the specification of the Hamiltonian constraint and its operator representation
have led to several quantization prescriptions. In this paper we have analyzed in
detail three of those most commonly used in the literature, known as the APS
\cite{aps-imp}, the sLQC \cite{acs}, and the MMO \cite{mmo-FRW} prescriptions. In
addition, we have introduced a new one, the sMMO prescription, which combines useful
features of both the MMO and the sLQC ones (see Sec.~\ref{sec:presc-sMMO}).

Basically, different prescriptions lead to slight differences in the evolution operator
$\hat{\Theta}$ that generates the unitary dynamical evolution in the internal time, whose
role is played by the massless scalar field. These differences make that the use of one or
another of the prescriptions results to be more convenient in distinct circumstances,
depending on the particular application under consideration.

In particular, the mathematical structure of the physical Hilbert space is different for
the various prescriptions discussed here. In fact, for generic superselection sectors, the
system has a rather more complicated structure in the APS and sLQC prescriptions, owing to
a twofold degeneracy of the spectrum of $\hat\Theta$, whereas in the same situations the
MMO and sMMO prescriptions (for which the spectrum is nondegenerate) provide a much
simpler Hilbert space of physical states. This fact has a significant influence in the
efficiency of the numerical techniques used in the dynamical study of the system, which
therefore varies considerably from the first to the second of these sets of prescriptions.
As discussed in Sec.~\ref{sec:numerics}, the construction and analysis of the physical
states in the degenerate cases requires more complicated methods, which in turn increase
the computational cost and the numerical error. Although this error is far from critical
in the computations that we have performed, since the relative error grows in the
degenerate cases, compared to the nondegenerate ones, from approximately $10^{-12}$ to
only $10^{-9}$, the problem of the time cost is relevant from the numerical point of view.
As discussed in Sec.~\ref{sec:num-basis}, the cost of the (most demanding) step, in which
the basis of $\Hil_{\phy}$ is constructed, is \emph{at least $8$ times} higher in
the degenerate cases than in the nondegenerate ones. This shows that, whenever the system
has to be analyzed numerically, the MMO and sMMO prescriptions are much more appropriate.
The cost difference becomes particularly critical once one tries to analyze more
complicated cosmological models, like for example Bianchi I \cite{hp-b1}.

Despite the significant focus on technical aspects, the main aim of our investigation has
been to identify and study possible differences between the considered prescriptions on a
physical level. To achieve this, we have analyzed a two-parameter family of physical
states with spectral profiles corresponding to a logarithmic normal distribution [see Eq.
\eqref{gauss-state}], and without imposing the restriction of semiclassicality. For our
analysis, we have chosen states peaked about low values of the scalar field momentum,
$\langle\hat{p}_{\phi}\rangle<500\hbar$, since the differences are easier to unveil in
this regime. For these states, we have been able to detect differences between the various
prescriptions by observing the interference pattern in the wave packet tail at the bounce.
The comparison of the states built in the different prescriptions, for the same spectral
profiles, has shown that the patterns are actually shifted with respect to each other.
This result confirms the existence of differences. Nonetheless, it does not allow one to
straightforwardly deduce which specific prescription has been employed, because the
commented shift depends also on other factors, such as the superselection sector and the
particular spectral profile of the state.

In order to address in depth the feasibility of the detection of differences between
prescriptions, further analysis has been performed. We have focused it on two fronts,
discussing the discrepancies in the measurements of standard cosmological observables, on
the one hand, and studying certain quantum operators which are specially sensible to a
change of prescription, on the other hand.

Concerning the first of these fronts, we have picked up three observables of interest in
cosmology, namely, the logarithmic volume, the Hubble parameter, and the scalar field
energy density. We have used them to compare the dynamical (quantum) trajectories of the
physical states specified above. We have evaluated the differences in the expectation
values of these observables between the different prescriptions, and shown that they are
several orders of magnitude smaller than the respective dispersions. As a consequence, and
as far as we restrict ourselves to these standard cosmological observables and to the
considered physical states, the differences between prescriptions are not detectable in
practice.

As for the other kind of observables that has been considered, we have computed the
expectation values of the operators $(\widehat{\Delta\Theta}_{\boldsymbol{AB}})^2$
defined in Eq. \eqref{eq:DeltaTheta}, which essentially encode the differences between the
Hamiltonian constraints that correspond to different prescriptions. These expectation
values are nonvanishing, reach the maximum at the bounce, and decay exponentially away
from it. The nature of the physical differences between prescriptions has been well
understood (see Sec.~\ref{sec:presc-diff}). The principal component from which these
differences arise comes from the potential term of $\hat\Theta$ in the volume momentum
representation. This potential term does not coincide for all the studied prescriptions.
While the subleading remnant in $\hat\Theta$ also varies when one changes the
prescription, this remnant is a compact operator and its effect is negligible. This also
explains why the smallest differences are observed between the sLQC and sMMO
prescriptions, since the potential term is the same in these two cases.

At this point, it is worth recalling that the studied physical trajectories and the
measured differences are genuinely well defined only if the spatial homogeneous slices are
compact (in the considered model, of $T^3$ topology). In noncompact cases, it is important
to take the limit in which the infrared regulator (the fiducial cell) is removed. This
step affects the observed difference. Indeed, taking states that correspond to the same
universe but with different fiducial cells, one can see that the effect of the compact
remnant gets removed once the cell $\cell$ tends to $\Sigma_t$. As a consequence, in the
limit when the regulator is removed, both the sLQC and the sMMO prescriptions can be
considered to converge in the physical sense discussed here (focusing the attention on the
kind of observables that we have introduced, constructed from $\hat\Theta$).

The existence of nontrivial differences shows that the prescriptions are truly physically
different, and the difference cannot be canceled out by a change of representation. This
fact has far going consequences, since, contrary to statements commonly found in the
literature \cite{cs-uniq-b1,*bt-eff}, the classical effective description of the system does
depend on the details of the quantization, and the characteristic effects of the
particular prescription that is used have to be taken into account in the process of
arriving to that description and determining its domain of validity.

\begin{acknowledgments}
  We would like to thank M.~Mart\'in-Benito and D.~Mart\'in-de Blas for discussions.
  This work was supported in part by the MICINN project FIS2008-06078-C03-03 and
  the Consolider-Ingenio program CPAN (CSD2007-00042) from Spain, and by the
  Natural Sciences and Engineering Research Council of Canada.
  T.P. acknowledges also the hospitality of the Institute of Theoretical Physics of
  Warsaw University and the financial support under grant of Minister Nauki i Szkolnictwa
  Wy{\.z}szego no.~N~N202~104838. J.O. acknowledges the support of CSIC under
  the grant No.~JAE-Pre~08~00791.
\end{acknowledgments}

\appendix

\section{WDW model}\label{app:WDW}

In this appendix, we describe a geometrodynamical analog of the system considered
in this paper.
This geometrodynamical model, built via a WDW quantization,
has been discussed extensively in the literature (see for example
Ref. \cite{aps-imp}). Here, we just summarize the properties necessary to define the
WDW limit of the LQC states.

The model is constructed following a process similar to that of the loop quantization (see
Sec.~\ref{sec:frame-quant}). The only difference is that now the geometry degrees
of freedom are quantized using a standard Schr\"odinger representation.
The kinematical Hilbert space is given again by a tensor product,
$\ub{\Hil}_{\kin}=\Hil^{\phi}\otimes\ub{\Hil}^{\gr}_{\kin}$, where $\Hil^{\phi}$ is the
space defined in Sec.~\ref{ssec:kin} and the gravitational Hilbert space is now
$\ub{\Hil}^{\gr}_{\kin}=L^2(\re,\rd v)$.
The triad operator still acts by multiplication, $\hat{p}|v)
= \textrm{sgn}(v)(2 \pi \gamma \lPl^2 \sqrt{\Delta}|v|)^{2/3}|v)$,
but the connection is now a well defined
derivative operator, $\hat{c} = 2i(2\pi\gamma\lPl^2)^{1/3}\Delta^{-1/3}|v|^{1/6}\partial_v|v|^{1/6}$, contrary to the
situation found in LQC.
Then, the evolution operator analog to $\hat{\Theta}$
(with a factor ordering compatible with that chosen for the latter operator) takes the form:
\begin{equation}
  \underline{\hat{\Theta}} = -12\pi G\sqrt{|v|}\partial_v|v|\partial_v\sqrt{|v|}.
\end{equation}
This operator is essentially selfadjoint in $\underline{\Hil}^{\rm grav}_{\kin}$. Its spectrum
is positive, twofold degenerate and absolutely continuous. Opposite orientations
of the triad ($v>0$ and $v<0$) are disjoint under the action of $\ub{\hat{\Theta}}$, therefore
the restriction to the (anti)symmetric sector can be implemented by considering only
the part $v>0$ and proceeding then to the (anti)symmetric completion of that part.
In the symmetric sector, there
exists an orthonormal basis of generalized eigenfunctions $(\underline{e}_k |$
of $\ub{\hat{\Theta}}$ whose elements are the rescaled
plane waves
\begin{equation}
  \ub{e}_{\pm k}(v)=(\ub{e}_{\pm k} |v)=\frac{1}{\sqrt{2\pi v}}e^{\pm ik\ln v} ,\quad v\in \mathbb{R}^+ .
\end{equation}
The corresponding eigenvalues are $\omega^2 = 12\pi G k^2$.
These generalized eigenfunctions satisfy the normalization condition
\begin{equation}\label{wdw:eig-norm}
  (\underline{e}_k | \underline{e}_{k'})=\delta(k-k').
\end{equation}
The group averaging procedure is straightforward to apply in this case, and provides the
Hilbert space of physical states $\ub{\Hil}_{\phy}=L^2(\re,\rd k)\ni\tilde{\ub{\Psi}}$, where
\begin{equation}
  \ub{\Psi}(v,\phi)=\int_\re \rd k \tilde{\ub{\Psi}}(k)\ub{e}_k(v) e^{i\omega(k)\phi}
\end{equation}
and $\omega(k)=\sqrt{12\pi G}|k|$.

\bibliography{mop-nongauss-va}{}
\bibliographystyle{apsrev4-1}

\end{document}